\documentclass[12pt]{article}

\usepackage[dvips]{graphicx}
\usepackage{amsmath}
\usepackage{amssymb}

\begin{document}

\begin{center}

\vspace{0.5cm}
\textbf{\Large Precise Coulomb wave functions \\ for a wide range of complex $\ell$, $\eta$ and $z$}

\vspace{5mm} {\large N.~Michel \footnote{\textit{E-mail address:} nmichel@utk.edu \\ \hspace*{0.5cm} \textit{Phone : 1-865-576-4295} \\
              \hspace*{0.5cm} \textit{Fax : 1-865-576-8746}}}

\vspace{3mm}
{\it
Department of Physics and Astronomy, University
of Tennessee, Knoxville, TN 37996, USA  \\
\vspace{1mm}
Physics Division,  Oak Ridge National
Laboratory, P.O.B.
2008, Oak Ridge, TN 37831, USA  \\
\vspace{1mm} Joint Institute for Heavy Ion Research, Oak Ridge, TN 37831, USA
}

\end{center}

\vspace{5mm}

\hrule \vspace{5mm} \noindent{\Large \bf Abstract}
\vspace*{5mm}

A new algorithm to calculate Coulomb wave functions with all of its arguments complex
is proposed. For that purpose, standard methods such as continued fractions and power/asymptotic series
are combined with direct integrations of the Schr{\"o}dinger equation in order to provide very stable calculations,
even for large values of $|\eta|$ or $|\Im(\ell)|$.
Moreover, a simple analytic continuation for $\Re(z) < 0$ is introduced, so that this zone of the complex $z$-plane does not pose any problem.
This code is particularly well suited for low-energy calculations and the calculation of resonances with extremely small widths.
Numerical instabilities appear, however, when both $|\eta|$ and $|\Im(\ell)|$ are large and $|\Re(\ell)|$ comparable or smaller than $|\Im(\ell)|$.

\vspace{5mm} \noindent{\Large \bf Program Summary}

\bigskip\noindent{\it Title of the program:} cwfcomplex

\bigskip\noindent{\it Catalogue number:}

\bigskip\noindent{\it Program obtainable from:}
                      CPC Program Library, \
                      Queen's University of Belfast, N. Ireland

\bigskip\noindent{\it Program summary URL:}

\bigskip\noindent{\it Licensing provisions:} none

\bigskip\noindent{\it Computers on which the program has been tested:} DELL GX400

\bigskip\noindent{\it Operating systems:} Linux, Windows

\bigskip\noindent{\it Programming language used:} C++

\bigskip\noindent{\it Memory required to execute with typical data:} 

\bigskip\noindent{\it No.\ of bits in a word:} 64

\bigskip\noindent{\it No.\ of processors used:} 1

\bigskip\noindent{\it Has the code been vectorized?:} No

\bigskip\noindent{\it No.\ of bytes in distributed program, including test data, etc.:}

\bigskip\noindent{\it No.\ of lines in distributed program:} 2422

\bigskip\noindent{\it Nature of physical problem:}
The calculation of Coulomb wave functions with all of their
arguments complex is revisited. The new methods introduced
allow to greatly augment the range of accessible $\ell$, $\eta$, and $z$.

\bigskip\noindent{\it Method of solution:}
Power/asymptotic series and continued fractions are supplemented with direct integrations of the 
Coulomb Schr{\"o}dinger equation. Analytic continuation for $\Re(z) < 0$ is also precisely
computed using linear combinations of the functions provided by standard methods,
which do not follow the branch cut requirements demanded for Coulomb wave functions.

\bigskip\noindent{\it Restrictions on the complexity of the problem:}

\bigskip\noindent{\it Typical running time:} N/A

\bigskip\noindent{\it Unusual features of the program:} none



\bigskip

\vspace{5mm} \noindent{\Large \bf Long Write-up}
\vspace*{5mm}

\section{Introduction}

Coulomb wave functions are one of the most basic objects of particle theory.
They describe the behavior of a particle in a point-like Coulomb field, and thus appear in virtually all domains of quantum physics.
The correspondent dimensionless Coulomb Schr{\"o}dinger equation reads: 
\begin{eqnarray}
&&w''(z) = \left( \frac{\ell(\ell+1)}{z^2} + \frac{2 \eta}{z} - 1 \right) w(z) \label{cwf_eq}
\end{eqnarray}
where $w(z)$ is a Coulomb wave function, $\ell$ is the orbital angular momentum, and $\eta$ the Sommerfeld parameter.

The Coulomb wave functions can be expressed with hyper-geometric functions \cite{Abramowitz_Coulomb}.
The regular Coulomb wave function reads:
\begin{eqnarray}
&&F_{\ell,\eta} (z) = C_\ell(\eta) \; z^{\ell+1} \; e^{i \omega z} \; _1F_1 \left( 1+\ell+i\omega\eta ; 2\ell+2 ; -2i \omega z \right) \label{F_analy} \\
&&C_\ell(\eta) = 2^\ell \; \exp \left[ \frac{-\pi \eta + \left[ \log (\Gamma(1+\ell+i\eta)) 
                                                              + \log(\Gamma (1+\ell-i\eta)) \right]}{2} - \log (\Gamma(2\ell+2)) \right] \label{Cl_eta}
\end{eqnarray}
In this expression, $\omega$ can be equal to $\pm 1$ and the normalizing Gamow factor $C_\ell(\eta)$ \cite{Abramowitz_Coulomb} is given.
Outgoing ($\omega = 1$) and incoming ($\omega = -1$) Coulomb wave functions are defined the following way:
\begin{eqnarray}
&&H^{\omega}_{\ell,\eta} (z)  = e^{i \omega \left[ z - \eta \log (2z) - \ell \frac{\pi}{2} + \sigma_\ell(\eta) \right]} 
\; _2F_0 \left( -\ell+i\omega\eta,1+\ell+i\omega\eta;;-\frac{i}{2 \omega z} \right) \label{H_analy} \\
&&\sigma_\ell(\eta) = \frac{\log(\Gamma (1+\ell+i\eta)) - \log (\Gamma(1+\ell-i\eta))}{2i} \label{sigma_l_eta}
\end{eqnarray}
where the Coulomb phase shift $\sigma_\ell(\eta)$ appears \cite{Abramowitz_Coulomb}.
The analytic continuation for complex $\ell$ and $\eta$ of Ref.~\cite{Humblet,Kolbig} for the function $\log(\Gamma(z))$ 
occurring in $C_\ell(\eta)$ and $\sigma_\ell(\eta)$ is followed, 
thus guaranteeing consistent values even when the negative real axis branch cut of complex variables $1+\ell+i\eta$ and $1+\ell-i\eta$ is crossed.
The regular Coulomb wave function $F_{\ell,\eta}$, as well as the logarithmic irregular Coulomb wave function $G_{\ell,\eta}$,
can be expressed with $H^{+}_{\ell,\eta}$ and $H^{-}_{\ell,\eta}$ \cite{Abramowitz_Coulomb}:
\begin{eqnarray}
&&F_{\ell,\eta} (z)  = \frac{H^{+}_{\ell,\eta} (z) - H^{-}_{\ell,\eta} (z)}{2i}  \label{F_from_H} \\
&&G_{\ell,\eta} (z)  = \frac{H^{+}_{\ell,\eta} (z) + H^{-}_{\ell,\eta} (z)}{2}  \label{G_from_H}
\end{eqnarray}

Despite the deceptively simple form of Eq.~(\ref{cwf_eq}) and analytical expressions of Coulomb wave functions of Eqs.~(\ref{F_analy},\ref{H_analy}),
the Coulomb wave function is difficult to compute numerically. Already on the real axis, it can vary by many orders of magnitude for moderate values of $|\eta|$.
The situation becomes even worse when the wave function is analytically continued to the complex plane. 
Analytic continuation arises when one deals, for example, with resonant states, as energies become complex \cite{Gamow}.
It appears also with non-integer values of $\ell$ with, for example, Regge pole trajectory calculations \cite{Regge_pole_ex}.
Coulomb wave functions are multivalued functions of the complex variable $z$ in the general case 
and thus a branch cut must be imposed on the negative $z$-real axis \cite{Humblet}. This implies that numerical calculations
must be employed with care, as wave functions issued directly from standard numerical expressions do not follow the same branch cut
discontinuities as the requested Coulomb wave function.

The Coulomb wave function computation has been considered in many papers. 
A recent review of numerical methods and definitions for both
non-relativistic and relativistic cases can be found in Ref.~\cite{Seaton}.
Most of papers have dealt with only real arguments \cite{Barnett_real}, or with
at least one of them real ($\ell$ in Ref.~\cite{Tamura}, $z$ in Ref.~\cite{Takemasa}).
The special important case of Whittaker functions, with $\eta$ purely imaginary 
(bound Coulomb wave functions), has also been treated on its own \cite{Noble}. The first paper (and only one to our knowledge) 
which considered all complex arguments in a unified way is Ref.~\cite{Thompson}. Through the use of continued fractions
calculated with the powerful Lentz method, as well as recurrence relations in $\ell$ and Pad{\'e} approximants, the authors managed to encompass
a large part of the complex plane for each $\ell$, $\eta$, and $z$. The program of Ref.~\cite{Thompson} 
quickly became a standard in the physics community and is part of the CERNLIB
library \cite{Cernlib}. However, important parts of the complex plane remained uncovered for both numerical and theoretical reasons, as was already stated
in \cite{Thompson}. First, because of numerical instabilities of used recursions, one cannot calculate Coulomb wave functions, for example, close to imaginary axes
when the modulus of $\eta$ or $\Im(\ell)$ becomes large. As a consequence, the case of large $|\eta|$ and integer $\ell$ for all $z$, 
important for low-energy narrow resonant states, has remained problematic \cite{Kruppa_comm}. 
Moreover, due to the different branch cuts of Coulomb wave functions and asymptotic series/continued
fractions in the complex $z$-plane, it is impossible to directly calculate $H^{\omega}_{\ell,\eta}(z)$ for $\Re(z) < 0$ and $\omega \Im(z) < 0$.

In order to circumvent these caveats, it has been chosen to complement standard methods with direct integrations of Eq.~(\ref{cwf_eq}).
The latter can be simply implemented, and the only requirement is that one has to integrate in directions of increasing modulus of the
wave function to avoid numerical instability \cite{Numerical_Recipes}. Also, the $\Re(z) < 0$ and $\omega \Im(z) < 0$ 
parts of the complex $z$-plane can be accessed with numerical methods, 
as the (wrong) Coulomb wave functions coming out of the latter are linear combinations of the true Coulomb wave functions
whose coefficients can be computed precisely, so that their determination becomes straightforward.
With these new features, it will be demonstrated that the range of arguments is much larger
than in previous programs. It is also important to state that quadruple precision is not needed with the proposed method.

The structure of the paper is as follows: first, the used numerical methods will be described in Sec.~(\ref{num}). 
Examples of calculations will then be depicted for several sets of arguments in Sec.~(\ref{ex}). In particular, 
the determination of resonant states with extremely small widths will be discussed. The structure of the program will then be described.
Finally, conclusions and perspectives will be stated.

\section {Numerical methods} \label{num}

\subsection {Power series for $F_{\ell,\eta}$} \label{F_PS_num}
The regular solution $F_{\ell,\eta}$ can be expanded in power series \cite{Abramowitz_Coulomb}:
\begin{eqnarray}
&&F_{\ell,\eta} (z) = C_\ell(\eta) \; \sum_{n=0}^{+\infty} a_n \; z^{n+\ell+1} \label{F_power_series} \\
&&a_0 = 1 \nonumber \\ 
&&a_1 = \frac{\eta}{\ell+1} \nonumber \\
&&a_n = \frac{2 \eta \; a_{n-1} - a_{n-2}}{n(n+2\ell+1)} \;\;\; \forall n \geq 2
\end{eqnarray}
This formula is very useful for small values of $|z|$, but is unstable for large $|z|$ because of numerical cancellations.
Hence, it is used only for $\displaystyle |z| \leq \frac{1}{2}$.

\subsection {Asymptotic series}
$H^{\omega}_{\ell,\eta}$ can be expanded in asymptotic series, so that for $|z|$ large enough, $H^{\omega}_{\ell,\eta}(z)$ can be calculated
up to a given numerical precision with a finite number of terms $N$ \cite{Todd}:
\begin{eqnarray}
&&H^{\omega}_{\ell,\eta} (z) \simeq e^{i\omega \left[ z - \eta \log (2z) - \ell \frac{\pi}{2} + \sigma_\ell(\eta) \right]}
                             \sum_{n=0}^{N-1} \frac{b_n}{z^{n}} \label{H_from_AS}\\
&&b_0 = 1 \nonumber \\
&&b_{n+1} = \frac{n(n+1 + 2i\omega\eta) + i\eta(i\eta + \omega) - \ell(\ell+1)}{2i\omega(n+1)} \; b_n \;\;\; \forall n \geq 0
\end{eqnarray}

However, due to the different branch cuts discontinuity of $H^{\omega}_{\ell,\eta}$ and the asymptotic series, Eq.~(\ref{H_from_AS})
is correct only for $\Re(z) > 0$ or $\Re(z) < 0$ and $\omega \Im(z) > 0$ \cite{Thompson}. 
In the rest of the complex plane, it nevertheless provides a linear combination of $H^{+}_{\ell,\eta}$ and $H^{-}_{\ell,\eta}$, which is utilized
to determine $H^{\omega}_{\ell,\eta}$ (see Sec.~(\ref{analy_cont_Re_lower_zero})).

In practice, the asymptotic series give a meaningful result for a given $N$ if $|a_N/z^N| < \epsilon$ with $\epsilon$ the numerical precision \cite{Todd}.
In addition, one checks if the Wronskian of the two functions generated by Eq.~(\ref{H_from_AS}) respectively using $\omega = 1$ and $\omega = -1$ 
is equal to $2i$. The Wronskian value can be evaluated with Eq.~(\ref{H_from_AS}) for $|z| \rightarrow +\infty$.
If $\Re(z) > 0$, $F_{\ell,\eta}(z)$ is calculated with asymptotic series using Eq.~(\ref{F_from_H}) 
if $H^{+}_{\ell,\eta}(z)$ and $H^{-}_{\ell,\eta}(z)$ are correctly computed with Eq.~(\ref{H_from_AS}).

\subsection {Continued fractions} \label{cont_frac}

The logarithmic derivatives $f^{\omega} = F_{\ell,\eta}' / F_{\ell,\eta}$
and $h^{\omega} = H^{\omega'}_{\ell,\eta} / H^{\omega}_{\ell,\eta}$ can be expanded in continued fractions \cite{Thompson}:
\begin{eqnarray}
&&f^{\omega}(z) = \frac{\ell+1}{z} + i\omega + \frac{1}{z} \left[ \frac{-2i\omega a z}     {b + 2i \omega z + } \mbox{ }
                                                           \frac{-2i\omega(a + 1) z}{b + 1 + 2i \omega z + \cdots} \mbox{ } \right] \label{f_cont_frac} \\
&&h^{\omega}(z) = i \omega \left( 1 - \frac{\eta}{z} \right) + \frac{i \omega}{z} \left[
                                                                            \frac{a c}{2(z - \eta + i\omega)+} \mbox{ }
                                                                            \frac{(a+1)(c + 1)}{2(z - \eta + 2i\omega) + \cdots} \right] \label{h_cont_frac}
\end{eqnarray}
where the standard notations $a = 1 + \ell + i\omega \eta$, $b = 2\ell + 2$, and $c = -\ell+i\omega\eta$ are used \cite{Thompson}.
The value of $f^{\omega}$ (also denoted as $f$) is derived from Eq.~(\ref{F_analy}) and is thus theoretically independent of $\omega$.
Lentz method is used to evaluate continued fractions numerically \cite{Thompson}.

The $f^{\omega}$ domain of convergence is the whole complex plane besides $F_{\ell,\eta}$ zeros, while the one of $h^{\omega}$ follows $_2F_0$ analytic properties,
so that it is the whole complex plane minus the half-axis $[0 : -i\omega\infty[$, where $h^{\omega}$ has a branch cut discontinuity.

The continued fraction $h^{\omega}$ is particularly important, as with the knowledge of $F_{\ell,\eta}(z)$, $F'_{\ell,\eta}(z)$,
and the Wronskian relation $F'_{\ell,\eta} H^{\omega}_{\ell,\eta} - F_{\ell,\eta} H^{\omega'}_{\ell,\eta} = 1$,
it can be used to determine $H^{\omega}_{\ell,\eta}(z)$ \cite{Thompson}:
\begin{eqnarray}
&&H^{\omega}_{\ell,\eta}(z) = \frac{1}{F_{\ell,\eta}(z) \left[ f(z) - h^{\omega}(z) \right]} \label{H_from_CF}
\end{eqnarray}
Note that $H^{\omega}_{\ell,\eta}(z)$ and $F_{\ell,\eta}(z)$ must be numerically linearly independent for this formula to be stable.
If they are not, $H^{-\omega}_{\ell,\eta}(z)$ is calculated instead and $H^{\omega}_{\ell,\eta}(z)$ can be deduced from it and Eq.~(\ref{F_from_H}).
As $h^{\omega}$ and $H^{\omega}_{\ell,\eta}$ have different branch cuts, $h^{\omega}$ is equal to the logarithmic derivative of $H^{\omega}_{\ell,\eta}$
only if $\Re(z) > 0$ or $\Re(z) < 0$ and $\omega \Im(z) > 0$, so that Eq.~(\ref{H_from_CF}) is correct in this zone only. 
However, the continued fraction can be used even outside this zone if one takes care of branch cuts (see Sec.~(\ref{analy_cont_Re_lower_zero})).
Added to that, the continued fractions $f^{\omega}$ and $h^{\omega}$ play a prominent role 
for the calculation of Coulomb wave functions by direct integration (see Sec.~(\ref{direct_int})).

The numerical applicability of these continued fractions is, however, hindered by spurious effects. It has been noticed in Ref.~\cite{Gautschi}
that Eq.~(\ref{f_cont_frac}) exhibits anomalous convergence. When $|z|$ becomes large, the general term of $f^{\omega}(z)$ becomes very small
before increasing very much, and then only to decrease again to provide a convergent result. 
As a consequence, both values of $f^{+}(z)$ and $f^{-}(z)$ are always calculated and compared to check convergence. 
However, the anomalous convergence phenomenon is weaker when one chooses $\omega$ such that $\omega \Im(z) < 0$ \cite{Gautschi},
so that $f^{\omega}(z)$ can be correct even if $f^{+}(z) \neq f^{-}(z)$ numerically.

The case of $h^{\omega}$ is much better \cite{Thompson}, but problems have nevertheless been encountered.
For example, the numerical value of $h^{+}(z)$ for $\ell = 0$, $\eta = 10$, and $z = 0.01 - 3i$ is wrong and numerically equal to $f(z)$.
This difficulty is removed by using only the $\omega$ for which $|f(z) - h^{\omega}(z)|$ is large enough 
(i.e. larger than 1 or at least larger than $|f(z) - h^{-\omega}(z)|$).

Another problem is the very slow convergence of Eq.~({\ref{h_cont_frac}}) in the vicinity of the branch cut for moderate $|z|$ (see Table (\ref{table_slow_cv})).
Direct integration is used to solve this problem.
For that, if the number of iterations in Lentz method exceeds 100,000 (one also assumes $\Re(z) \geq 0$), one calculates $H^{\omega}_{\ell,\eta}(z_0)$
and $H^{\omega'}_{\ell,\eta}(z_0)$ with $z_0$ not too close to the imaginary axis, 
chosen so that $|H^{\omega}_{\ell,\eta}|$ increases from $z_0$ to $z$.
The slow convergence of $h^{\omega}$ is absent for $z_0$, so that Eq.~(\ref{h_cont_frac}) can be used for the integration starting point.
Then, one integrates Eq.~(\ref{cwf_eq}) from $z_0$ to $z$, 
which is a stable operation as $|H^{\omega}_{\ell,\eta}|$ increases along the integration path. 
$h^{\omega}(z)$ is then equal to $H^{\omega'}_{\ell,\eta}(z)/H^{\omega}_{\ell,\eta}(z)$ at the end of integration. If $\Re(z) < 0$,
one uses the symmetry formula $h^{\omega}_{\ell,\eta}(z) = -h^{-\omega}_{\ell,-\eta}(-z)$. This formula can be demonstrated using the fact that
$H^{\omega}_{\ell,\eta}(z) \propto H^{-\omega}_{\ell,-\eta}(-z)$ for $\omega \Im(z) > 0$ 
(both functions are solutions of Eq.~(\ref{cwf_eq}) and are minimal in the considered region for $|z| \rightarrow +\infty$), and analytic continuation 
as both functions $f_1 : z \rightarrow h^{\omega}_{\ell,\eta}(z)$ and $f_2 : z \rightarrow -h^{-\omega}_{\ell,-\eta}(-z)$ have the same branch cut.

\subsection {Direct integration} \label{direct_int}
Considering the simplicity of the Coulomb equation (Eq.~(\ref{cwf_eq})), direct integration is a suitable method to calculate Coulomb wave functions.
For that, the Burlisch-Stoer-Henrici method of Ref.~\cite{Numerical_Recipes} is used.
However, one has to pay attention to two problems. Firstly, no branch cut discontinuity can come out of direct integration, so that it is necessary
to integrate in the zones of the complex plane where branch cut effects are absent. Hence, numerical integration is performed only for $\Re(z) > 0$.
For the other half of the complex plane, one uses the symmetry transformation $z \rightarrow -z$, $\eta \rightarrow -\eta$, leaving Eq.~(\ref{cwf_eq}) invariant.
Secondly, numerical integration is stable only if the modulus of the Coulomb wave function increases or remains close to constant.
Increase or decrease of the wave function along the integration path is determined by its second-order Taylor expansion at $z = z_0+h$:
\begin{eqnarray}
&&\frac{\Psi(z)}{\Psi(z_0)} \simeq 1 + h\frac{\Psi'(z_0)}{\Psi(z_0)} + \frac{h^2}{2} \left( \frac{\ell(\ell+1)}{z_0^2} + \frac{2\eta}{z_0} - 1 \right)
\label{Taylor_exp}
\end{eqnarray}
where $\Psi$ is either $F_{\ell,\eta}$ or $H^{\omega}_{\ell,\eta}$, $h$ the integration step,
and ($z_0$, $\Psi(z_0)$, $\Psi'(z_0)$) the starting point of the numerical integration.
If the modulus of the ratio defined in Eq.~(\ref{Taylor_exp}) is larger than one, the numerical integration can be performed safely. 
If not, the continued fraction $q(z) = \Psi'(z)/\Psi(z)$ is evaluated with Eq.~(\ref{f_cont_frac}) ($q=f^{\omega}$)
or Eq.~(\ref{h_cont_frac}) ($q = h^{\omega}$). Eq.~(\ref{cwf_eq}) is integrated backward from $z$ to $z_0$ 
with ($z$, 1, $q(z)$) as the starting point, guaranteeing stable integration. One then obtains ($z_0$, $\Psi_c(z_0)$, $\Psi'_c(z_0)$) after integration,
with obviously $\Psi_c(z_0) = \Psi(z_0)/\Psi(z)$ and $\Psi'_c(z_0) = \Psi'(z_0)/\Psi(z)$. The value of ($z$, $\Psi(z)$, $\Psi'(z)$) comes forward.
The only nuisance in this method is that the continued fraction $q(z)$ can be wrong due to numerical instability (see Sec.~(\ref{cont_frac})).

This can be partially solved if one considers a direct integration of $F_{\ell,\eta}(z)$. 
If $\Re(\ell) > -1$, $F_{\ell,\eta}$ increases in modulus with $|z|$ in the vicinity of $z=0$ (non-oscillatory zone).
As a consequence, if $|F_{\ell,\eta}|$ is found to decrease on its initial path,
$z_0$ is reinitialized to $z/(2|z|)$, where the power series formula of Eq.~(\ref{F_power_series}) is available,
so that a decrease of $|F_{\ell,\eta}|$ from $z_0$ to $z$ is less likely to happen. 
If the direct integration of $F_{\ell,\eta}(z)$ is found to be unstable despite this change of path, it is preferred
to calculate $H^{\omega}_{\ell,\eta}(z)$ with direct integration ($\omega$ chosen so the branch cut of $h^{\omega}_{\ell,\eta}$ is avoided) 
as $h^{\omega}$ is numerically more stable than $f^{\omega}$ (see Sec.~(\ref{cont_frac})).
$F_{\ell,\eta}(z)$ is then calculated with Eq.~(\ref{F_from_H}) and the following formula: 
\begin{eqnarray}
&&H^{-\omega}_{\ell,\eta}(z) = \frac{2i\omega}{H^{\omega}_{\ell,\eta}(z) \left[ h^{\omega}(z) - h^{-\omega}(z) \right]} \label{H_min_omega_from_CF}
\end{eqnarray}
which can be obtained similarly to Eq.~(\ref{H_from_CF}).
This process is, however, stable if $H^{\omega}_{\ell,\eta}(z)$ and $H^{-\omega}_{\ell,\eta}(z)$ are not numerically equal, so that it is not employed
if $|F_{\ell,\eta}|$ is found to be smaller than 0.1 on its integration path. This is a sound procedure as $f^{\omega}(z)$ is usually correct
in this case if $\omega \Im(z) < 0$ (see Sec.~(\ref{cont_frac})).

As a result, direct integration is a powerful tool to determine $F_{\ell,\eta}(z)$ outside the zone of applicability of Eq.~(\ref{F_power_series}).
It also provides $h^{\omega}(z)$ close to its branch cut, where the numerical cost of Eq.~(\ref{h_cont_frac}) becomes prohibitive
(see Sec.~(\ref{cont_frac})).

\subsection {$H^{\omega}_{\ell,\eta}$ expansion}
When $2\ell$ is not an integer, the following formula can be used to calculate $H^{\omega}_{\ell,\eta}$ \cite{Thompson}:
\begin{eqnarray}
&&H^{\omega}_{\ell,\eta} = \frac{F_{\ell,\eta} \; e^{i \omega \chi} - F_{-\ell-1,\eta}}{\sin{\chi}} \label{Homega_exp} \\
&&\chi = \sigma_\ell(\eta) - \sigma_{-\ell-1} (\eta) - (\ell+1/2)\pi \label{chi}
\end{eqnarray}

In practice, it has been chosen to apply it only for $|\Im(\ell)| \geq 1$ and $|z| \leq 1$, as other methods have been found
to be more robust for other cases. 
For a given $z$, the expression of Eq.~(\ref{Homega_exp}) is numerically stable 
if the Wronskian relation between $F_{\ell,\eta}$ and $F_{-\ell-1,\eta}$ is respected: 
\begin{eqnarray}
&&F'_{\ell,\eta} F_{-\ell-1,\eta} - F_{\ell,\eta} F'_{-\ell-1,\eta} = \sin{\chi} \label{wronskian_chi}
\end{eqnarray}

Note that Eq.~(\ref{chi}) is not used to calculate $\sin{\chi}$ as it is unstable due to cancellation effects.
Another formula is preferred:
\begin{eqnarray}
&&\sin{\chi} = -(2\ell+1) \; C_\ell(\eta) \; C_{-\ell-1}(\eta) \label{sin_chi}
\end{eqnarray}
which can be demonstrated using Eqs.~(\ref{F_power_series},\ref{wronskian_chi}) with $z \rightarrow 0$.

\subsection {Analytic continuation for $\Re(z) < 0$} \label{analy_cont_Re_lower_zero}

Analytic continuation for $\Re(z) < 0$ is first considered for the regular function $F_{\ell,\eta}$.
Using Eqs.~(\ref{Cl_eta},\ref{F_power_series}), $F_{\ell,\eta}(z)$ and $F_{\ell,-\eta}(-z)$ can be shown to be proportional:
\begin{eqnarray}
&&F_{\ell,\eta}(z) = - e^{-\pi \left( \eta - i \ell \right)} \; F_{\ell,-\eta}(-z) \mbox{ for } \arg (z) > 0    \nonumber \\
&&F_{\ell,\eta}(z) = - e^{-\pi \left( \eta + i \ell \right)} \; F_{\ell,-\eta}(-z) \mbox{ for } \arg (z) \leq 0 \label {analy_cont_F}
\end{eqnarray}
Hence, $F_{\ell,\eta}(z)$ can always be deduced from $F_{\ell,-\eta}(-z)$, so that calculations for $\Re(z) \geq 0$ are sufficient to determine
$F_{\ell,\eta}$ in all the complex plane.

The situation is more complicated for $H^{\omega}_{\ell,\eta}$, as the direct evaluation of Eq.~(\ref{H_from_AS}) and Eq.~(\ref{H_from_CF})
provides correct values for $\Re(z) < 0$ and $\omega \Im(z) > 0$, but wrong results occur when $\Re(z) < 0$ and $\omega \Im(z) < 0$ \cite{Thompson}.
We will denote as $H^{\omega \mbox{ } (AS_{d})}_{\ell,\eta}$ and $H^{\omega \mbox{ } (CF_{d})}_{\ell,\eta}$ the numerical values coming 
from a naive implementation of respectively Eq.~(\ref{H_from_AS}) and Eq.~(\ref{H_from_CF}). 
As they are issued from analytic expressions providing solutions of Eq.~(\ref{cwf_eq}),
they are still solutions of this equation even when $\Re(z) < 0$ and $\omega \Im(z) < 0$. However, the different branch cuts of
$H^{\omega}_{\ell,\eta}$, $H^{\omega \mbox{ } (AS_{d})}_{\ell,\eta}$, and $H^{\omega \mbox{ } (CF_{d})}_{\ell,\eta}$ imply that the two latter functions
are linear combinations of $H^{+}_{\ell,\eta}$ and $H^{-}_{\ell,\eta}$ in this quadrant of the $z$-complex plane.
Their coefficients will be shown to be very simple expressions of $\ell$ and $\eta$ and can be related to standard circuital relations \cite{Dzieciol}.

If one considers $|z| \rightarrow +\infty$, one can deduce from Eqs.~(\ref{H_analy},\ref{H_from_AS}) that:
\begin{eqnarray}
&&H^{\omega}_{\ell,\eta}(z) = H^{\omega \mbox{ } (AS_{d})}_{\ell,\eta}(z) + a_{\omega} \; H^{-\omega \mbox{ } (AS_{d})}_{\ell,\eta}(z)
\mbox{ for } \Re(z) < 0 \mbox{ and }\omega \Im(z) < 0 \label{H_omega_linear_combination}
\end{eqnarray}
where $a_{\omega}$ is a constant depending on $\ell$ and $\eta$. 
The equality $H^{-\omega}_{\ell,\eta}$ = $H^{-\omega \mbox{ } (AS_{d})}_{\ell,\eta}$ in the considered region is also used.
From Eqs.~(\ref{F_from_H},\ref{H_omega_linear_combination}), one has:
\begin{eqnarray}
&&2i F_{\ell,\eta}(x_-) = H^{+ \mbox{ } (AS_{d})}_{\ell,\eta}(x_-) + (a_+ - 1) \; H^{- \mbox{ } (AS_{d})}_{\ell,\eta}(x_-) 
\label{a_plus_eq} \\
&&-2i F_{\ell,\eta}(x_+) = H^{- \mbox{ } (AS_{d})}_{\ell,\eta}(x_+) + (a_- - 1) \; H^{+ \mbox{ } (AS_{d})}_{\ell,\eta}(x_+)
\label{a_minus_eq}
\end{eqnarray}
where $x_+ = x + i \epsilon$ and $x_- = x - i \epsilon$ for $x < 0$ and $\epsilon > 0$.
Branch cut discontinuities of $F_{\ell,\eta}$ and $H^{\omega \mbox{ } (AS_{d})}_{\ell,\eta}$ are straightforward from Eqs.~(\ref{F_power_series},\ref{H_from_AS}),
so that Eq.~(\ref{a_plus_eq}) can be rewritten as:
\begin{eqnarray}
&&2i \; e^{-2 i \pi \ell} \; F_{\ell,\eta}(x_+)  = e^{-2 \pi \eta} \; H^{+ \mbox{ } (AS_{d})}_{\ell,\eta}(x_+) 
+ e^{2 \pi \eta} \; (a_+ - 1) H^{- \mbox{ } (AS_{d})}_{\ell,\eta}(x_+) + O(\epsilon) 
\label{a_plus_eq_rewritten}
\end{eqnarray}
Finally, using Eqs.~(\ref{a_minus_eq},\ref{a_plus_eq_rewritten}) 
with $\epsilon \rightarrow 0$, one obtains $a_{\omega} = 1 - e^{2i\pi \left( i \eta - \ell \omega \right)}$ and hence the requested formula:
\begin{eqnarray}
&&H^{\omega}_{\ell,\eta}(z) = H^{\omega \mbox{ } (AS_{d})}_{\ell,\eta}(z)
+ \left[ 1 - e^{2i\pi \left( i \eta - \ell \omega \right)} \right] H^{-\omega \mbox{ } (AS_{d})}_{\ell,\eta}(z) \label{analy_cont_H_AS}
\end{eqnarray}
for which $\Re(z) < 0$ and $\omega \Im(z) < 0$. 

Considering the different branch cut discontinuities of Eqs.~(\ref{H_from_AS},\ref{H_from_CF}) on the negative real axis and analytic continuation, one obtains:
\begin{eqnarray}
&&H^{\omega \mbox{ } (CF_{d})}_{\ell,\eta}(z) = e^{2i\pi \left( \ell \omega - i \eta \right)} H^{\omega \mbox{ } (AS_{d})}_{\ell,\eta}(z)
\label{AS_CF_connection}
\end{eqnarray}
with $\Re(z) < 0$ and $\omega \Im(z) < 0$.

Using Eqs.~(\ref{F_from_H},\ref{analy_cont_H_AS},\ref{AS_CF_connection}), the formulas analog to Eq.~(\ref{analy_cont_H_AS}) for continued fractions are derived:
\begin{eqnarray}
&&H^{\omega}_{\ell,\eta}(z) = H^{\omega \mbox{ } (CF_{d})}_{\ell,\eta}(z) 
                            - 2 i \omega \left[ e^{2i\pi \left( \ell \omega - i \eta \right)} - 1 \right] F_{\ell,\eta}(z)
\label{analy_cont_H_CFa} \\
&&H^{\omega}_{\ell,\eta}(z) = H^{-\omega \mbox{ } (CF_{d})}_{\ell,\eta}(z) 
                            + 2 i \omega \; e^{-2i\pi \left( \ell \omega + i \eta \right)} \; F_{\ell,\eta}(z)
\label{analy_cont_H_CFb}
\end{eqnarray}
with $\Re(z) < 0$, $\omega \Im(z) < 0$ for Eq.~(\ref{analy_cont_H_CFa}) but $\omega \Im(z) > 0$ for Eq.~(\ref{analy_cont_H_CFb}).
As the calculation of $F_{\ell,\eta}(z)$ is prerequisite to determine $H^{\omega}_{\ell,\eta}(z)$ with continued fraction formulas (see Eq.~(\ref{H_from_CF})),
the numerical evaluation of $H^{\omega}_{\ell,\eta}(z)$ with Eqs.~(\ref{analy_cont_H_CFa},\ref{analy_cont_H_CFb}) is straightforward.

Even though the expressions of the coefficients in front of $H^{-\omega}_{\ell,\eta}$ in Eq.~(\ref{analy_cont_H_AS})
and $F_{\ell,\eta}$ in Eqs.~(\ref{analy_cont_H_CFa},\ref{analy_cont_H_CFb}) are elementary, 
care must be given to calculate them due to possible overflow or underflow and numerical cancellations.
One has to use complex generalizations of the standard C-language functions $log1p(x) = \log(1+x)$ and $expm1(x) = e^x - 1$ 
for $x \rightarrow 0$ to avoid possible numerical inaccuracies.

As a result, the $\Re(z) < 0$ domain no longer poses any numerical problem, 
as the formulas of Eqs.~(\ref{analy_cont_H_AS},\ref{analy_cont_H_CFa},\ref{analy_cont_H_CFb}) render it comparable to the rest of the $z$-complex plane.

\subsection{Poles of the Coulomb wave functions}
When $1+l+i \omega \eta$ is a negative integer, Coulomb wave functions are undefined (see Eqs.~(\ref{Cl_eta},\ref{sigma_l_eta})). 
Nevertheless, if one considers $\Re(\ell) > -1$, numerical solutions of Eq.~(\ref{cwf_eq})
always exist and can be computed. For this, $F_{\ell,\eta}$ is defined with Eq.~(\ref{F_power_series}) putting arbitrarily $C_\ell(\eta) = 1$.
Direct integration can be performed precisely for $F_{\ell,\eta}$, as the continued fraction $f^{\omega}$ of Eq.~(\ref{f_cont_frac}) is finite so that
no numerical inaccuracy can occur. $H^{-\omega}_{\ell,\eta}$ can still be defined with Eq.~(\ref{H_from_CF}) as $h^{-\omega} \neq f^{\omega}$, 
so that one can calculate two linearly independent solutions of Eq.~(\ref{cwf_eq}) when $1+l+i \omega \eta$ is a negative integer.
Note that the branch cut of $H^{-\omega}_{\ell,\eta}$ is, for this definition, $[0:i \omega \infty[$ and not the negative real axis. 
$H^{\omega}_{\ell,\eta}$ and $G_{\ell,\eta}$ can, however, not be defined so that they are arbitrarily put equal to
$F_{\ell,\eta}$ and $H^{-\omega}_{\ell,\eta}$ respectively.

\subsection {Quasi-real $\ell$, $\eta$ and $z$} \label{first_order_method}
When $\ell$, $\eta$ or $z$ are very close to their real axes with at least one of them complex, the imaginary part
of Coulomb wave functions can become tens of order of magnitude smaller than their real parts.
Consequently, it can be numerically imprecise as calculations are always provided up to the same absolute precision for both real and imaginary parts.
This is especially visible if one deals with resonant states of extremely small widths $\gamma$ 
such as proton emitters ($\gamma \sim 10^{-20}$ keV) \cite{Kruppa_PRL}.
For these kinds of states, one generally uses approximate current formulas \cite{Kruppa_proc}, providing very good values for $\gamma$ if it is small enough.
However, it is possible to reach the same precision directly at the Coulomb wave function level. For this, one expands the 
Coulomb wave functions $F_{\ell,\eta}$ and $G_{\ell,\eta}$ up to first order in the vicinity of the real axes of $\ell$, $\eta$, and $z$:
\begin{eqnarray}
A(\ell,\eta,z) &=& A(\ell_r,\eta_r,x) + i \left[ \ell_i \frac{\partial A}{\partial \ell} (\ell_r,\eta_r,x)
                                               + \eta_i \frac{\partial A}{\partial \eta} (\ell_r,\eta_r,x)
                                               + y      \frac{\partial A}{\partial z} (\ell_r,\eta_r,x) \right] \nonumber \\
&+& O(\ell_i^2,\eta_i^2,y^2) \label{FG_exp}
\end{eqnarray}
where $\ell_r,\eta_r,x$ and $\ell_i,\eta_i,y$ are respectively the real and imaginary parts of $\ell$, $\eta$, and $z$, 
and $A(\ell,\eta,z)$ is either $F_{\ell,\eta}(z)$ or $G_{\ell,\eta}(z)$.
All values involving $A(\ell_r,\eta_r,x)$ are real (one considers $x > 0$ only) so that function and partial derivatives can be evaluated numerically. 
In practice, if $\epsilon$ is the demanded numerical precision, 
the conditions $|y| < \sqrt{\epsilon} \min(1,x)$, $|\eta_i| < \sqrt{\epsilon}$  and $|\ell_i| < \sqrt{\epsilon}$ must be fulfilled for Eq.~(\ref{FG_exp})
to be used.
It was checked that direct and approximate results yield the same results up to a precision comparable to $\epsilon$
if $|y| = \sqrt{\epsilon} \min(1,x)$, $|\ell_i| = |\eta_i| = \sqrt{\epsilon}$ with $\epsilon = 10^{-10}$ (see Table (\ref{tab_first_order_exp})).

$H^{\omega}_{\ell,\eta}(z)$ can be obtained straightforwardly from the knowledge of $F_{\ell,\eta}(z)$ and $G_{\ell,\eta}(z)$.
Consequently, the whole Coulomb wave function can be derived up to a given relative numerical precision for both real and imaginary parts 
even if they are very different in modulus.

\subsection{Scaled wave functions and alternative normalization}
It often happens that $H^{\omega}_{\ell,\eta}(z)$ overflows or underflows when $|z|$ or $|\eta|$ become large, but only through the exponential factor 
of Eq.~(\ref{H_analy}). As a consequence, the following scaled Coulomb wave functions can also be calculated in the code:
\begin{eqnarray}
&&H^{\omega}_{\ell,\eta}(z)_{sc} = H^{\omega}_{\ell,\eta}(z) \; e^{-i \omega \left[ z - \eta \log(2z) \right]} \label{H_scaled}
\end{eqnarray}
They are particularly useful if one calculates products of Coulomb wave functions where the different exponential factors cancel each other.

At the limit of very small energies, where $|\eta|$ is very large, $C_\ell(\eta)$ can also overflow or underflow, so that it is no longer possible
to calculate Coulomb wave functions. However, their normalization factor is usually unimportant, as in the case of a resonant state calculation.
For that, we introduced the following renormalized wave functions:
\begin{eqnarray}
&&F_{\ell,\eta}(z)^r = C_\ell(\eta)^{-1} \; F_{\ell,\eta}(z)  \nonumber \\
&&H^{\omega}_{\ell,\eta}(z)^r = C_\ell(\eta) \; H^{\omega}_{\ell,\eta}(z)  \nonumber \\
&&G_{\ell,\eta}(z)^r = C_\ell(\eta) \; G_{\ell,\eta}(z) \label{renorm_functions}
\end{eqnarray}
An example of a resonant state for which $C_\ell(\eta)$ underflows will be given in Sec.~(\ref{proton_states}).

\subsection{Recurrence relations and associated Wronskian tests} \label{rec_rel}
Coulomb wave functions obey recurrence relations of their angular momentum $\ell$ \cite{Thompson}:
\begin{eqnarray}
&&w_{\ell,\eta}(z) = \frac{S_\ell}{R_\ell} \; w_{\ell-1,\eta}(z) - \frac{1}{R_\ell} \; w'_{\ell-1,\eta}(z) \nonumber \\
&&w'_{\ell,\eta}(z) = R_\ell \; w_{\ell-1,\eta}(z) - S_\ell \; w_{\ell,\eta}(z) \label{rec_rel_eq}
\end{eqnarray}
where $w$ is any of the $F,G,H^{+}$ or $H^{-}$ functions, $\displaystyle R_\ell = (2\ell + 1) \frac{C_\ell(\eta)}{C_{\ell-1}(\eta)}$
and $\displaystyle S_\ell = \frac{\ell}{z} + \frac{\eta}{\ell}$. $\Re(\ell)$, denoted $\ell_r$, is supposed to be larger than zero.
These recurrence relations are stable provided $|w_{\ell,\eta}(z)|$ increases with $\ell_r$.
As $F_{\ell,\eta}(z) \rightarrow 0$ with $\ell_r \rightarrow +\infty$ \cite{Thompson}, for $\ell_r$ large enough,
the recurrence relations are stable with $\ell_r$ decreasing if one calculates regular Coulomb wave functions
and with $\ell_r$ increasing if one calculates irregular Coulomb wave functions. For the irregular wave functions, one has the most
stable calculations if one calculates $H^{\omega}_{\ell,\eta}(z)$ with $\omega$ chosen so 
$|H^{\omega}_{\ell_0,\eta}(z)| \leq |H^{-\omega}_{\ell_0,\eta}(z)|$, $\ell_0$ being the angular momentum of smallest modulus.
Indeed, this guarantees $H^{\omega}_{\ell_0,\eta}$ to be the minimal solution of Eq.~(\ref{cwf_eq}) if $F_{\ell_0,\eta}$ is not.
One may have, however, a turning point $\ell_t$ before which $|F_{\ell,\eta}(z)|$ increases \cite{Thompson}. In this case, $F_{\ell,\eta}$ has to be recurred
backward from the angular momentum of largest modulus, denoted $\ell_1$, backward to $\ell_t$ but forward from $\ell_0$ to $\ell_t$. 
Conversely, $H^{\omega}_{\ell,\eta}$ must be recurred backward from $\ell_t$ to $\ell_0$ and forward from $\ell_t$ to $\ell_1$.

The previous recurrence relations provide additional relations between Coulomb wave functions:
\begin{eqnarray}
&&F_{\ell,\eta} H^{\omega}_{\ell+1,\eta} - F_{\ell+1,\eta} H^{\omega}_{\ell,\eta} = \frac{1}{R_{\ell+1}} \label{l_wronskian} \\
&&F'_{\ell,\eta} H^{\omega}_{\ell+1,\eta} + F_{\ell,\eta} H^{\omega'}_{\ell+1,\eta} 
= F'_{\ell+1,\eta} H^{\omega}_{\ell,\eta} + F_{\ell+1,\eta} H^{\omega'}_{\ell,\eta}  \label{l_wronskian_der}
\end{eqnarray}
If $\omega$ is chosen so $|H^{\omega}_{\ell,\eta}(z)| \leq |H^{-\omega}_{\ell,\eta}(z)|$, 
in order to have two Coulomb wave functions numerically linearly independent,
Eqs.~(\ref{l_wronskian},\ref{l_wronskian_der}) 
provide a good test to check the accuracy of the Coulomb wave functions calculated with the methods of previous sections,
as angular momentum recurrence relations do not enter them.

The non-standard normalization defined in Eq.~(\ref{renorm_functions}) can also be used in the code along with recurrence relations, 
which can be obtained straightforwardly from Eqs.~(\ref{renorm_functions},\ref{rec_rel_eq}). 
Scaling of $H_{\ell,\eta}^{\omega}$ for $\omega = \pm1$ (see Eq.~(\ref{H_scaled}))) is, however, not considered in this context, 
because all Coulomb wave functions have to be numerically finite for the method to work, so that it would only result in the trivial multiplication
of $H_{\ell,\eta}^{\omega}(z)$ by $e^{-i \omega \left[ z - \eta \log(2z) \right]}$.

\section{Examples} \label{ex}

\subsection{Calculations in difficult zones of the complex plane} \label{log_cwf_mod_calc}

\begin{center}
\begin{figure}[hbt]
\includegraphics[width=9cm,angle=-90]{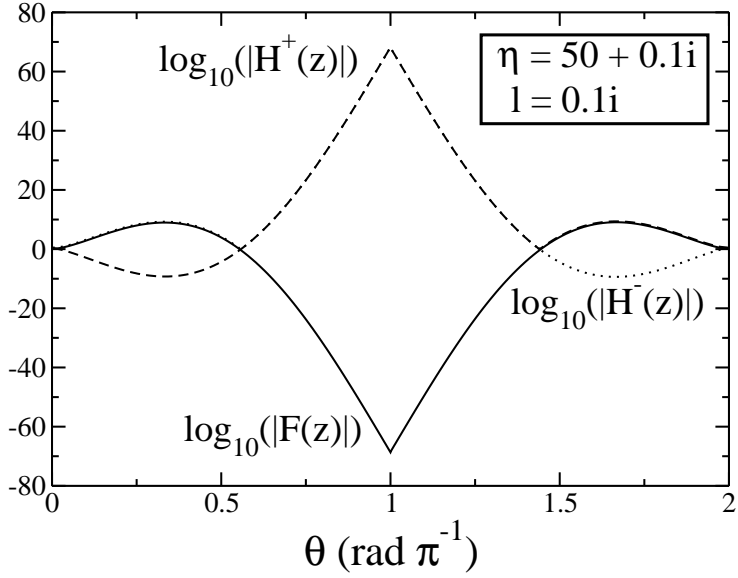}
\caption{Decimal logarithms of $|F_{\ell,\eta}(z)|$, $|H^{+}_{\ell,\eta}(z)|$, and $|H^{-}_{\ell,\eta}(z)|$
as a function of the argument $\theta$ of $z$, given in units of radian over $\pi$.
They are respectively represented with full, dashed, and dotted lines.
$z = R_t e^{i \theta}$ with $R_t$ the generalized turning point $|\eta| + \sqrt{|\ell(\ell+1)| + |\eta|^2}$.
$\ell$ and $\eta$ are here respectively equal to $0.1i$ and $50+0.1i$.}
\label{fig1}
\end{figure}
\end{center}

\begin{center}
\begin{figure}[hbt]
\includegraphics[width=9cm,angle=-90]{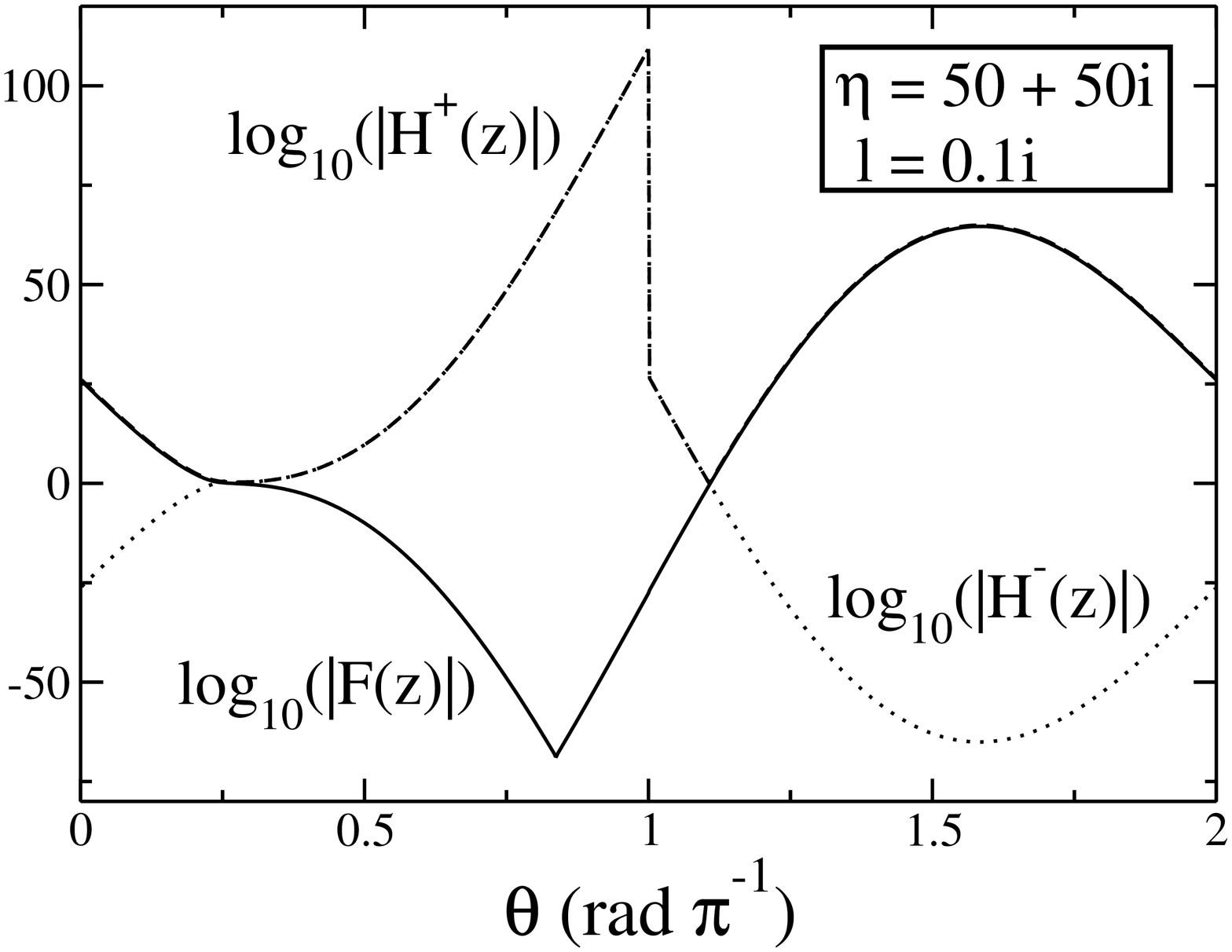}
\caption{Same as Fig.~(\ref{fig1}), except that $\ell$ and $\eta$ are here respectively equal to $0.1i$ and $50+50i$.}  
\label{fig2}
\end{figure}
\end{center}

\begin{center}
\begin{figure}[hbt]
\includegraphics[width=9cm,angle=-90]{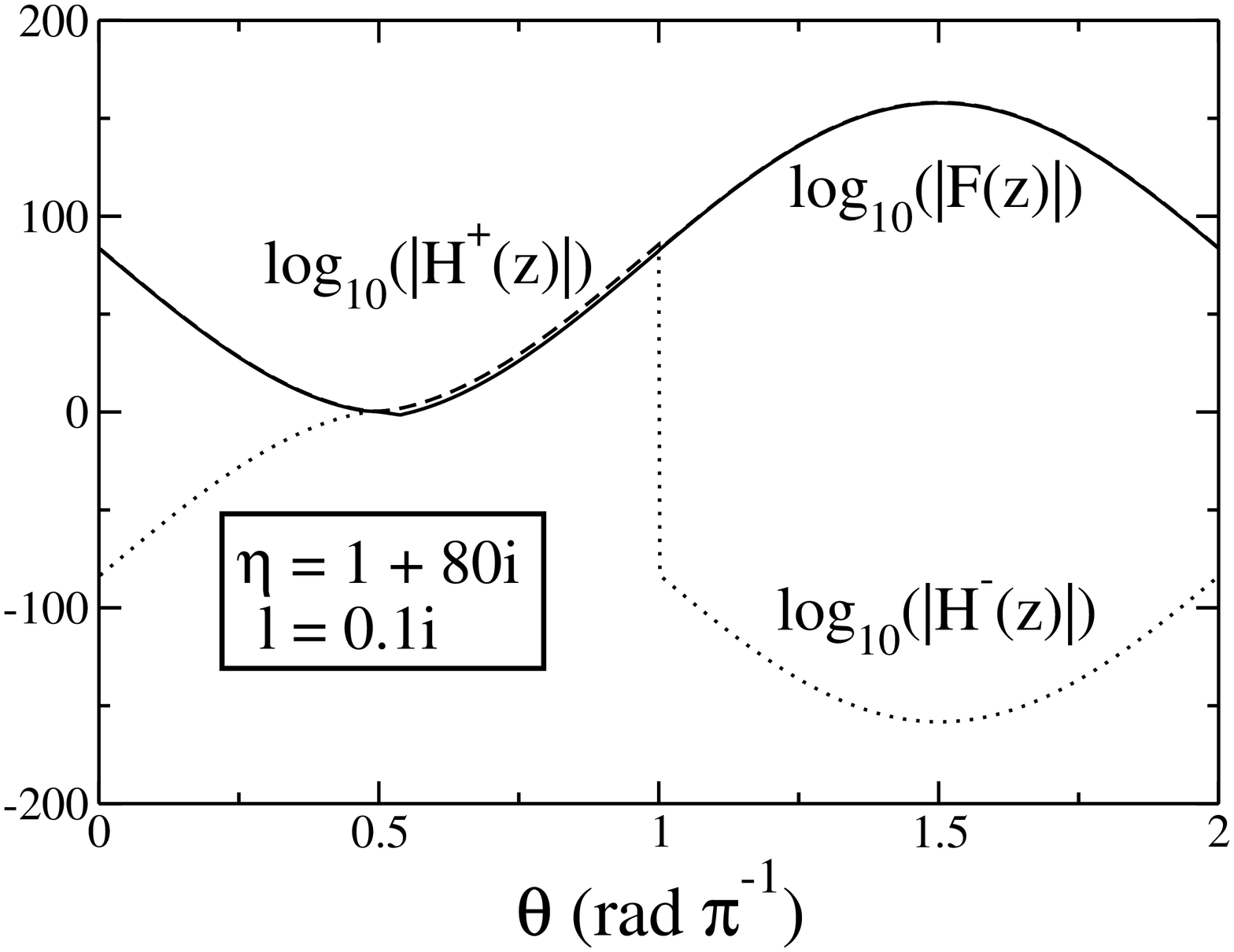} 
\caption{Same as Fig.~(\ref{fig1}), except that $\ell$ and $\eta$ are here respectively equal to $0.1i$ and $1+80i$.}
\label{fig3}
\end{figure}
\end{center}

\begin{center}
\begin{figure}[hbt]
\includegraphics[width=9cm,angle=-90]{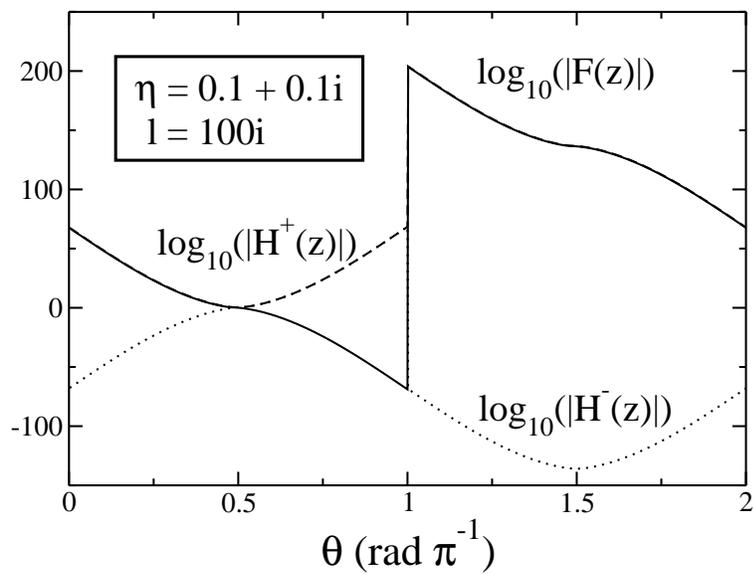}
\caption{Same as Fig.~(\ref{fig1}), except that $\ell$ and $\eta$ are here respectively equal to $100i$ and $0.1+0.1i$.}
\label{fig4}
\end{figure}
\end{center}

\begin{center}
\begin{figure}[hbt]
\includegraphics[width=9cm,angle=-90]{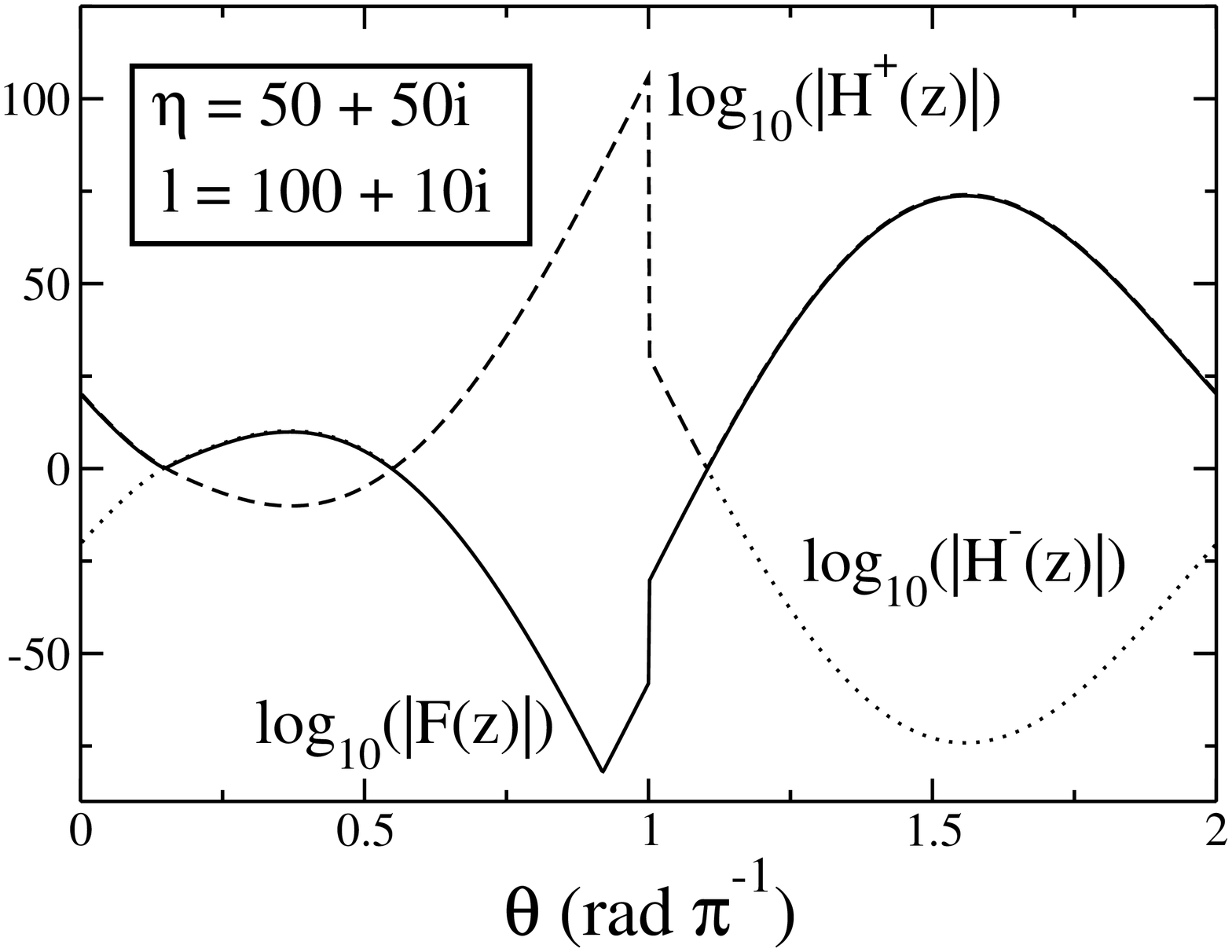}
\caption{Same as Fig.~(\ref{fig1}), except that $\ell$ and $\eta$ are here respectively equal to $100+10i$ and $50+50i$.}
\label{fig5}
\end{figure}
\end{center}

In order to illustrate the proposed numerical methods, we selected sets of $\ell$, $\eta$, and $z$ parameters having sizable values.
$z$ is always of the form $\displaystyle R_t e^{i \theta}$, where $\displaystyle R_t = |\eta| + \sqrt{|\ell(\ell+1)| + |\eta|^2}$ 
is a generalization of the turning point in the complex plane. Problems of convergence indeed typically occur in the vicinity of this point \cite{Thompson}.
$|\eta|$ and $|\ell|$ have been chosen so that at least one is large in a set of parameters
(see Figs.~(\ref{fig1},\ref{fig2},\ref{fig3},\ref{fig4},\ref{fig5})). 

It was chosen not to have both $|\eta|$ and $|\Im(\ell)|$ large with $|\Re(\ell)| < |\Im(\ell)|$ or comparable, as calculations become
unstable therein. For these values of $\ell$, $\eta$, and $z$, Coulomb wave functions vary by several orders of magnitude along the complex circle
of radius $R_t$ and oscillate much as well. Hence, we represented the decimal logarithm of the modulus of Coulomb wave functions, which varies smoothly.
One should note that the discontinuities encountered at $\theta = \pi$ follow the branch cuts imposed to Coulomb wave functions 
and are not induced by any numerical inaccuracy. These calculations show that Coulomb wave functions can be calculated precisely even when they vary
much in argument and modulus.

\subsection{Calculations of resonant states of very small widths} \label{proton_states}
In order to show the possibilities of the present program related to resonant states with extremely small widths, e.g.~proton emitters \cite{Kruppa_PRL}, 
we consider a spherical Schr{\"o}dinger equation with a Woods-Saxon potential crudely mimicking
a heavy nuclear target acting on a proton projectile:
\begin{eqnarray}
&&h = \frac{p^2}{2\mu} - \frac{V_0}{1 + \exp \left( \frac{r-R_0}{d} \right)} + V_c(r) \label{h_proton_emitter}
\end{eqnarray}
where $\mu$ is the reduced mass of the proton state so that $\hbar^2/2\mu = 21$ MeV fm$^{2}$,
$d$ is the diffuseness of the potential fixed at 0.63 fm, $R_0$ is the radius of the potential of 6.5 fm,
$V_0$ is the depth of the Woods-Saxon potential, and $V_c(r)$ is a Coulomb potential generated by a uniformly charged sphere of radius $R_0$ and charge $Z = 66$.
These parameters correspond to the proton emitter $^{141}$Ho \cite{Kruppa_proc}. 

The $2s_{1/2}$ proton state energy and width of the Hamiltonian $h$ of Eq.~(\ref{h_proton_emitter}) is calculated for several 
values of $V_0$ (see Table (\ref{table_proton_emitter})). The width $\gamma$ coming from direct integration of $h$ is compared
with the following standard current approximation \cite{Kruppa_proc}:
\begin{eqnarray}
&&\gamma_{c} =  \Re(k) \; \frac{\hbar^2}{\mu} \; \left| \frac{u(R)}{H^+(kR)} \right|^2 \label{gamma_app}
\end{eqnarray}
where $k$ is the linear momentum of the proton resonant state, $u(r)$ its radial wave function,
and $R$ a radius large enough so $u(r) \propto H^+(kr)$ for $r > R$. The value $R = 20$ fm was chosen.
As expected, both values $\gamma$ and $\gamma_c$ are identical, as $\gamma$ is very small.
If $V_0 = 56.46$ MeV, one obtains an energy of 6.949$\cdot 10^{-4}$ MeV while the width is numerically zero. This value
could be computed only through the renormalization of Coulomb wave functions of Eq.~(\ref{renorm_functions}), 
as for this energy $\log_{10}(C_\ell(\eta)) = -535$, implying $C_\ell(\eta)$ underflow.

These results show that the proposed program is very well suited for the direct calculation of very narrow resonances,
for which one has to enter numerically challenging areas of the complex plane. It would be interesting to use
this program along with coupled-channel integration methods \cite{Kruppa_PRL,Kruppa_proc} in order to extend the current method to deformed states.

\section{The program \textit{cwfcomplex}}

\subsection{Routines of the program}
The code \textit{cwfcomplex} is written in standard C++, uses only standard libraries and is thus portable on many machines.
It is separated in four different files: \textit{complex\_functions.H}, \textit{cwfcomp.H}, \textit{cwfcomp.cpp} and \textit{test\_rec\_rel.cpp}.

\textit{complex\_functions.H} contains elementary complex functions which are not in the standard library,
and routines calculating constants specific to the Coulomb wave functions:
\begin{itemize}
\item \textit{inf\_norm}: provides the infinite norm of a complex number.
\item \textit{isfinite}: returns true if the complex number is finite.
\item operators overloading of complex and integers.
\item \textit{expm1}: complex generalization of the function $expm1(x) = e^x - 1$ precise for $x \rightarrow 0$.
\item \textit{log1p}: complex generalization of the function $log1p(x) = \log (1+x)$ precise for $x \rightarrow 0$.
\item \textit{log\_Gamma}: $\log(\Gamma(z))$ calculated with the method of Ref.~\cite{Kolbig}.
\item \textit{sigma\_l\_calc}: complex Coulomb phase shift $\sigma_{\ell}(\eta)$ (see Eq.~(\ref{sigma_l_eta})).
\item \textit{log\_Cl\_eta\_calc}: $\log(C_{\ell}(\eta))$ (see Eq.~(\ref{Cl_eta})).
\item \textit{log\_cut\_constant\_AS\_calc}: logarithm of the constant in front of $H^{-\omega \mbox{ } (AS_{d})}_{\ell,\eta}$ in Eq.~(\ref{analy_cont_H_AS}).
\item \textit{log\_cut\_constant\_CFa\_calc}: logarithm of the constant in front of $F_{\ell,\eta}$ in Eq.~(\ref{analy_cont_H_CFa}).
\item \textit{log\_cut\_constant\_CFb\_calc}: logarithm of the constant in front of $F_{\ell,\eta}$ in Eq.~(\ref{analy_cont_H_CFb}).
\item \textit{sin\_chi\_calc}: $\sin(\chi)$ calculated with Eq.~(\ref{sin_chi}).
\item \textit{exp\_I\_omega\_chi\_calc}: $e^{i \omega \chi}$ calculated with Eqs.~(\ref{chi},\ref{sin_chi}).
\end{itemize}

In \textit{cwfcomp.H}, the class \textit{ODE\_integration} and member functions \textit{extrapolation\_in\_zero}, \textit{F\_r\_u},
\textit {integration\_Henrici} and \textit{operator ()} performing direct integration of the Coulomb Schr{\"o}dinger equation are defined, 
as well as the class \textit{Coulomb\_wave\_functions}, with which one can calculate all Coulomb wave functions.
All the routines of the class \textit{Coulomb\_wave\_functions} are in \textit{cwfcomp.cpp} .
\begin{itemize}
\item \textit{F\_dF\_init}: initialization of the member variables \textit{debut}, \textit{F\_debut} and \textit{dF\_debut} used for direct integration 
(see Sec.~(\ref{recommendations})).
\item \textit{asymptotic\_series}: calculate the asymptotic series in Eq.~(\ref{H_from_AS}) for Coulomb wave function and derivative. 
\item \textit{continued\_fraction\_f, continued\_fraction\_h}: calculate the continued fractions of respectively 
Eq.~(\ref{f_cont_frac}) and Eq.~(\ref{h_cont_frac}).
\item \textit{F\_dF\_power\_series}: calculate $F_{\ell,\eta}(z)$ and $F'_{\ell,\eta}(z)$ with Eq.~(\ref{F_power_series}).
\item \textit{asymptotic\_expansion\_F\_dF, asymptotic\_expansion\_H\_dH\_scaled}: calculate $F_{\ell,\eta}(z)$ and $F'_{\ell,\eta}(z)$ 
                                                               ($H^{\omega}_{\ell,\eta}(z)$ and $H^{\omega'}_{\ell,\eta}(z)$ scaled, see Eq.~(\ref{H_scaled}))
                                                                from Eqs.~(\ref{H_from_AS},\ref{analy_cont_H_AS},\ref{H_scaled}).
\item \textit{F\_dF\_direct\_integration, H\_dH\_direct\_integration}: calculate $F_{\ell,\eta}(z)$ and $F'_{\ell,\eta}(z)$ 
                                                    ($H^{\omega}_{\ell,\eta}(z)$ and $H^{\omega'}_{\ell,\eta}(z)$) by direct integration of Eq.~(\ref{cwf_eq}).
\item \textit{partial\_derivatives, first\_order\_expansions}: calculate $F_{\ell,\eta}(z)$ and $F'_{\ell,\eta}(z)$ or $G_{\ell,\eta}(z)$ and $G'_{\ell,\eta}(z)$
                                                            with the method of Sec.~(\ref{first_order_method}).
\item \textit{H\_dH\_from\_first\_order\_expansions}: calculate $H^{\omega}_{\ell,\eta}(z)$ and $H^{\omega'}_{\ell,\eta}(z)$ 
                                                  with the method of Sec.~(\ref{first_order_method}).
\item \textit{H\_dH\_with\_F\_dF\_and\_CF}: calculate $H^{\omega}_{\ell,\eta}(z)$ and $H^{\omega'}_{\ell,\eta}(z)$ 
                                       with Eqs.~(\ref{h_cont_frac},\ref{analy_cont_H_CFa},\ref{analy_cont_H_CFb}).
\item \textit{H\_dH\_with\_expansion}: calculate $H^{\omega}_{\ell,\eta}(z)$ and $H^{\omega'}_{\ell,\eta}(z)$ with Eq.~(\ref{Homega_exp}).
\item \textit{F\_dF\_with\_symmetry\_relations}: calculate $F_{\ell,\eta}(z)$ and $F'_{\ell,\eta}(z)$ for $\Re(z) < 0$ with Eq.~(\ref{analy_cont_F}).
\end{itemize}
Except \textit{F\_dF\_init}, all the latter routines are private in the class \textit{Coulomb\_wave\_functions} and should not be used directly.

The following routines of the class \textit{Coulomb\_wave\_functions} are public and provide the requested Coulomb wave functions:
\begin{itemize}
\item \textit{F\_dF}: calculate $F_{\ell,\eta}(z)$ and $F'_{\ell,\eta}(z)$.
\item \textit{G\_dG}: calculate $G_{\ell,\eta}(z)$ and $G'_{\ell,\eta}(z)$.
\item \textit{H\_dH}: calculate $H^{\omega}_{\ell,\eta}(z)$ and $H^{\omega'}_{\ell,\eta}(z)$.
\item \textit{H\_dH\_scaled}: calculate $H^{\omega}_{\ell,\eta}(z)$ and $H^{\omega'}_{\ell,\eta}(z)$ scaled (see Eq.~(\ref{H_scaled})). 
\end{itemize}
The calculation of $G_{\ell,\eta}(z)$ and $G'_{\ell,\eta}(z)$ is performed by calculating $F_{\ell,\eta}(z)$, $H^{+}_{\ell,\eta}(z)$,
and their derivatives, so that one has $G_{\ell,\eta}(z) = H^{+}_{\ell,\eta}(z) - i F_{\ell,\eta}(z)$ and 
$G'_{\ell,\eta}(z) = H^{+'}_{\ell,\eta}(z) - i F'_{\ell,\eta}(z)$.

The file \textit{test\_rec\_rel.cpp} contains additional useful routines using the class \textit{Coulomb\_wave\_functions}:
\begin{itemize}
\item \textit{Wronskian\_test}: function calculating Coulomb wave functions accuracy from Wronskian's (see Sec.~(\ref{rec_rel})).
\item \textit{cwf\_l\_tables\_recurrence\_relations}: routine calculating Coulomb wave functions with recurrence relations for integer spaced $\ell$'s 
(see Sec.~(\ref{rec_rel})).
\item \textit{F\_dF\_l\_tables\_rec\_rel\_helper, cwf\_l\_tables\_rec\_rel\_helper}: 
routines called by \\ \textit{cwf\_l\_tables\_recurrence\_relations}, not intended to be used directly.
\end{itemize}

\subsection{Use of the program} \label{prog_use}
Due to its object-oriented programming, \textit{cwfcomplex} is easy to use. One has to declare first a class \textit{Coulomb\_wave\_functions}
with three parameters \textit{l}, \textit{eta}, and \textit{is\_it\_normalized}, where \textit{l} and \textit{eta} are two complex numbers 
representing $\ell$ and $\eta$, and \textit{is\_it\_normalized} is a boolean equal to \textit{true} 
if one uses the standard normalization of Coulomb wave functions and \textit{false} if one uses the normalization of Eq.~(\ref{renorm_functions}).
For example, one declares \textit{class Coulomb\_wave\_functions cwf(is\_it\_normalized,l,eta);}.
Then, one can use the member functions of the class \textit{cwf}. For that, one needs 
the complex value \textit{z}, the integer \textit{omega} $= \pm 1$ (for $H^{\omega}_{\ell,\eta}$),
and two complex numbers \textit{A, dA} to store the Coulomb wave function and its derivative.
The instructions to obtain \textit{A} and \textit{dA} are the following:  
\begin{itemize}
\item \textit{cwf.F\_dF (z,A,dA);} to compute $F_{\ell,\eta}(z)$ and $F'_{\ell,\eta}(z)$. 
\item \textit{cwf.G\_dG (z,A,dA);} to compute $G_{\ell,\eta}(z)$ and $G'_{\ell,\eta}(z)$. 
\item \textit{cwf.H\_dH (omega,z,A,dA);} to compute $H^{\omega}_{\ell,\eta}(z)$ and $H^{\omega'}_{\ell,\eta}(z)$.
\item \textit{cwf.H\_dH\_scaled (omega,z,A,dA);} to compute $H^{\omega}_{\ell,\eta}(z)$ and $H^{\omega'}_{\ell,\eta}(z)$ scaled (see Eq.~(\ref{H_scaled})).
\end{itemize}

In order to test the accuracy of previous functions, one has to declare a second class \textit{Coulomb\_wave\_functions} of parameters
\textit{is\_it\_normalized}, \textit{l+1} and \textit{eta}, for example, \\ \textit{class Coulomb\_wave\_functions cwf\_p(is\_it\_normalized,l+1,eta);}.
Then, the instruction \textit{const double W = Wronskian\_test (z,cwf,cwf\_p);} provides their relative precision, 
calculated from their Wronskian's and angular momentum recurrence relations (see Sec.~(\ref{rec_rel})), stored in the double \textit{W}.

Tables of integer spaced $\ell$'s Coulomb wave functions are calculated with the routine \\ \textit{cwf\_l\_tables\_recurrence\_relations}.
For this, one needs the complex angular momentum of smallest modulus \textit{l\_deb}, the number of angular momenta to calculate \textit{Nl},
\textit{eta}, \textit{is\_it\_normalized}, the number of $z$-variables \textit{Nz}, the one-dimensional complex array of $z$-variables \textit{z\_tab},
six two-dimensional \textit{Nz} x \textit{Nl} complex arrays \textit{F\_tab}, \textit{dF\_tab}, \textit{G\_tab}, \textit{dG\_tab}, 
\textit{Hp\_tab}, \textit{dHp\_tab}, \textit{Hm\_tab} and \textit{dHm\_tab} 
to respectively store the regular wave functions $F$ and derivatives, the irregular wave functions $G$ and derivatives,
the outgoing irregular wave functions $H^+$ and derivatives, and the incoming irregular wave function $H^-$ and derivatives, so that 
\textit{F\_tab[iz][il]} will provide $F_{\ell,\eta}(z)$ with $\ell=$\textit{l\_deb+il} and $z=$\textit{z\_tab[iz]} (same for other tables).
The instruction is then: \\ \\
\textit{cwf\_l\_tables\_recurrence\_relations (l\_deb,Nl,eta,is\_it\_normalized,Nz,z\_tab,} \\
\hspace*{5.8cm} \textit{F\_tab,dF\_tab,G\_tab,dG\_tab,Hp\_tab,dHp\_tab,Hm\_tab,dHm\_tab);} \\ \\
If one considers a single complex variable \textit{z}, one can use the following instruction: \\ \\
\textit{cwf\_l\_tables\_recurrence\_relations (l\_deb,Nl,eta,is\_it\_normalized,z,} \\
\hspace*{5.8cm} \textit{F\_tab,dF\_tab,G\_tab,dG\_tab,Hp\_tab,dHp\_tab,Hm\_tab,dHm\_tab);} \\ \\
where \textit{F\_tab,...,dHm\_tab} are now one-dimensional arrays of \textit{Nl} complex numbers 
(\textit{F\_tab[il]} = $F_{\ell,\eta}(z)$ with $\ell=$\textit{l\_deb+il}, same for other tables).

Examples are provided by the program \textit{examples.cpp}, which calculates values of different 
Coulomb wave functions and derivatives on a circular $z$-path of the form $z = R e^{i \theta}$, with $\theta \in [0:2 \pi[$,
for given $\ell_0$, $\eta$ and $R$, their accuracy with the function \textit{Wronskian\_test} and a table of integer spaced $\ell$'s
Coulomb wave functions with the routine \textit{cwf\_l\_tables\_recurrence\_relations}, with the same parameters as before and $\ell$ starting from $\ell_0$.

\subsection{Recommendations} \label{recommendations}
Even though there are no restrictions for the complex values used in the program, it is advised to use $\Re(\ell) > -1$.
Calculations have indeed been found to be more stable for these values.
If one has $\Re(\ell) \leq -1$, one can use the symmetry transformation $\ell \rightarrow -\ell-1$, as Eqs.~(\ref{Homega_exp},\ref{chi})
imply $H^{\omega}_{\ell,\eta} = H^{\omega}_{-\ell-1,\eta} \; e^{i \omega \chi}$.

Due to the direct integration procedures, Coulomb wave functions should be calculated if possible for $z$ varying smoothly 
if one considers tables of Coulomb wave function values.
Indeed, the complex numbers $z,F_{\ell,\eta}(z),F'_{\ell,\eta}(z)$ are stored in the class under the names \textit{debut}, \textit{F\_debut} and \textit{dF\_debut}
after each calculation, so that the integration from this point to the next is faster and more precise if $z$ varies continuously in the complex plane.
Also, it is better for $|F_{\ell,\eta}|$ to increase on its path as then no continued fraction calculation $f(z)$ 
(see Eq.(\ref{f_cont_frac})) is needed during direct integration.

\section{Conclusion}
The computation of Coulomb wave functions with all its arguments complex is a very difficult task.
The single use of power/asymptotic series and continued fractions quickly shows its limitation when $|\eta|$ or $|\Im(\ell)|$ increases. 
It was found that the range of accessible $\ell$, $\eta$, and $z$ is greatly augmented by adding the direct integration method of the Coulomb equation.
Calculations are stable for values of $|\Im(\eta)|$ as important as 80, and $|\Im(\ell)|$ can be as large as 100 as well. 
This method is particularly stable for the implementation of extremely narrow resonant states.
However, instabilities appear when both $|\eta|$ and $|\Im(\ell)|$ are large, and $|\Re(\ell)|$ smaller or comparable to $|\Im(\ell)|$. 
For example, the values $\ell = 15i$, $\eta = 10$, and $z = 20 e^{i\theta}$
used in a calculation similar to the ones presented in Sec.~(\ref{log_cwf_mod_calc}) provide wrong wave functions in the vicinity 
of $\theta = 3 \pi / 2$. This particular problem can be treated by always accepting the value of $f^{\omega}(z)$ in \textit{F\_dF\_direct\_integration}
for backward integration (see Sec.~(\ref{direct_int})).
Other issues can be solved by using \textit{H\_dH\_with\_expansion} in \textit{H\_dH} and \textit{H\_dH\_scaled} even if $|z| > 1$ or $|\Im(\ell)| < 1$.
(The comments in the code beginning with four slashes explain to the user how to make modifications accordingly.)
Nevertheless, these are solutions for very particular cases and cannot be included in a general program.
Calculations can also become too long if one considers irregular Coulomb wave functions for $0 < |\ell| < 1$ and $|z| < 10^{-5}$,
as the continued fraction of Eq.(\ref{h_cont_frac}) converges very slowly for $z \sim 0$ \cite{Thompson} and direct integration cannot be used in this region.
Even though one encounters numerical problems for large values of $|\Im(\ell)/\Re(\ell)|$ and $|\eta|$ or very small $|z|$, this program has rendered
possible calculations which could not be undertaken with standard methods.

\section*{Acknowledgments}
Discussions with A.T.~Kruppa and J.~Rotureau are gratefully acknowledged.
This work was supported in part by the U.S.~Department of Energy
under Contracts Nos.~DE-FG02-96ER40963 (University of Tennessee),
DE-AC05-00OR22725 with UT-Battelle, LLC (Oak Ridge National
Laboratory), and DE-FG05-87ER40361 (Joint Institute for Heavy Ion Research).

\newpage
\vspace{5mm} \noindent{\Large \bf Test Input} \\
{ \\
true \\
(1,0.1) \\
(50,50) \\
100.156 \\
10 \\
3 \\ \\
\# Description of the input parameters \\
\# \\ 
\# Boolean: true if ones uses standard normalization, false if one uses alternative normalization. \\
\# Complex: angular momentum l. \\
\# Complex: Sommerfeld parameter eta. \\
\# Double:  radius R of the path in the complex plane: z = R exp(i theta), theta in [0:2 pi[. \\
\# Integer: number of points Nz to be considered on the path. \\
\# Integer: number of points Nl for the recurrence relation : l[rec] = l+k, k in [0:Nl-1]. \\
\# \\
\# Compilation: g++ -O3 examples.cpp -o run \\
\# Input instruction: ./run $<$ test.input \\
\# The output is in test.output .
}
\newpage

\vspace{5mm} \noindent{\Large \bf Test Output}
{\\ \\ is\_it\_normalized:true l:(1,0.1) eta:(50,50) R:100.156  Nz:10 Nl:3 \\ \\
z:(100.156,0)\\
F:(-1.021072923e+15,-2.836755456e+15) F':(1.275057299e+15,-2.729507771e+15) \\
 G:(2.836755456e+15,-1.021072923e+15) G':(2.729507771e+15,1.275057299e+15) \\
 H+:(5.673510913e+15,-2.042145845e+15) H+':(5.459015542e+15,2.550114598e+15) \\
 H-:(7.0774288e-17,1.501204734e-16) H-':(5.671783379e-17,-1.558437769e-16) \\
 Wronskian test: 1.628444125e-11\\ \\
z:(81.02790609,58.87021973) \\
 F:(0.01090170509,0.002924757522) F':(0.006665318369,0.003695114571) \\
 G:(57.24722492,-32.54791917) G':(-42.75162529,10.97359983) \\
 H+:(57.24430017,-32.53701746) H+':(-42.7553204,10.98026514) \\
 H-:(57.25014968,-32.55882087) H-':(-42.74793017,10.96693451) \\
 Wronskian test: 1.307402399e-11\\ \\
z:(30.94990609,95.25401645) \\
 F:(-2.246133078e-15,2.098754042e-15) F':(-5.747597654e-16,2.506104287e-15) \\
 G:(-4.367342675e+13,-1.907186698e+14) G':(1.181536987e+14,1.103266317e+14) \\
 H+:(-4.367342675e+13,-1.907186698e+14) H+':(1.181536987e+14,1.103266317e+14) \\
 H-:(-4.367342675e+13,-1.907186698e+14) H-':(1.181536987e+14,1.103266317e+14) \\
 Wronskian test: 2.604836552e-11\\ \\
z:(-30.94990609,95.25401645) \\
 F:(-3.696304706e-35,8.374503306e-35) F':(5.116568262e-35,9.162544125e-35) \\
 G:(2.32622983e+33,-4.170545023e+33) G':(2.19801873e+33,4.986604576e+33) \\
 H+:(2.32622983e+33,-4.170545023e+33) H+':(2.19801873e+33,4.986604576e+33) \\
 H-:(2.32622983e+33,-4.170545023e+33) H-':(2.19801873e+33,4.986604576e+33) \\
 Wronskian test: 6.182547672e-12\\ \\
z:(-81.02790609,58.87021973) \\
 F:(-2.432130956e-67,3.004725207e-66) F':(3.593950134e-66,1.98955822e-66) \\
 G:(1.065320986e+65,-5.911485321e+64) G':(1.320210869e+64,1.651750319e+65) \\
 H+:(1.065320986e+65,-5.911485321e+64) H+':(1.320210869e+64,1.651750319e+65) \\
 H-:(1.065320986e+65,-5.911485321e+64) H-':(1.320210869e+64,1.651750319e+65) \\
 Wronskian test: 1.818126864e-10\\ \\
z:(-100.156,1.22651674e-14) \\
 F:(7.915510206e-34,-4.070932761e-34) F':(3.191449042e-34,1.291346723e-33) \\
 G:(4.180561145e+103,8.128671328e+103) G':(-1.326122108e+104,3.277393326e+103) \\
 H+:(4.180561145e+103,8.128671328e+103) H+':(-1.326122108e+104,3.277393326e+103) \\
 H-:(4.180561145e+103,8.128671328e+103) H-':(-1.326122108e+104,3.277393326e+103) \\
 Wronskian test: 1.24249569e-16\\ \\
z:(-81.02790609,-58.87021973) \\
 F:(-23318.74764,-17080.04412) F':(28164.502,-34900.64756) \\
 G:(17080.04412,-23318.74763) G':(34900.64758,28164.50199) \\
 H+:(34160.08824,-46637.49527) H+':(69801.29514,56329.004) \\
 H-:(6.998646405e-06,8.686284079e-06) H-':(1.393963282e-05,-1.022684069e-05) \\
 Wronskian test: 1.320762168e-11\\ \\z:(-30.94990609,-95.25401645) \\
 F:(-3.419604636e+30,-3.206140946e+30
) F':(4.148673182e+30,-5.870853625e+30) \\
 G:(3.206140946e+30,-3.419604636e+30) G':(5.870853625e+30,4.148673182e+30) \\
 H+:(6.412281891e+30,-6.839209271e+30) H+':(1.174170725e+31,8.297346365e+30) \\
 H-:(4.01631585e-32,5.686566419e-32) H-':(7.772167842e-32,-7.290658234e-32) \\
 Wronskian test: 1.127990924e-11\\ \\
z:(30.94990609,-95.25401645) \\
 F:(1.125583254e+40,3.548477279e+39) F':(3.759922307e+38,1.698605313e+40) \\
 G:(-3.548477279e+39,1.125583254e+40) G':(-1.698605313e+40,3.759922307e+38) \\
 H+:(-7.096954559e+39,2.251166509e+40) H+':(-3.397210626e+40,7.519844613e+38) \\
 H-:(6.373392552e-43,-2.945348841e-41) H-':(-4.036846683e-41,1.270436304e-41) \\
 Wronskian test: 1.056406572e-11\\ \\
z:(81.02790609,-58.87021973) \\
 F:(-2.579395538e+32,7.380968215e+32) F':(-9.701448492e+32,1.926734513e+32) \\
 G:(-7.380968215e+32,-2.579395538e+32) G':(-1.926734513e+32,-9.701448492e+32) \\
 H+:(-1.476193643e+33,-5.158791075e+32) H+':(-3.853469026e+32,-1.940289698e+33) \\
 H-:(-4.963907579e-34,-9.779175601e-35) H-':(2.098665903e-34,6.035174252e-34) \\
 Wronskian test: 1.628966764e-12\\ \\ \\
Recurrence relations results for a table of z values.\\
z:(100.156,0) \\
 l[rec]:(1,0.1) \\
 F:(-1.021072923e+15,-2.836755456e+15) F':(1.275057299e+15,-2.729507771e+15) \\
 G:(2.836755456e+15,-1.021072923e+15) G':(2.729507771e+15,1.275057299e+15) \\
 H+:(5.673510913e+15,-2.042145845e+15) H+':(5.459015542e+15,2.550114597e+15) \\
 H-:(7.0774288e-17,1.501204734e-16) H-':(5.67178338e-17,-1.558437769e-16)\\ \\l[rec]:(2,0.1) \\
 F:(-9.963131598e+14,-2.756374147e+15) F':(1.235449857e+15,-2.655381893e+15) \\
 G:(2.756374147e+15,-9.963131598e+14) G':(2.655381893e+15,1.235449857e+15) \\
 H+:(5.512748294e+15,-1.99262632e+15) H+':(5.310763785e+15,2.470899715e+15) \\
 H-:(7.256451117e-17,1.54534022e-16) H-':(5.855852326e-17,-1.602324976e-16)\\ \\l[rec]:(3,0.1) \\
 F:(-9.584372325e+14,-2.64073598e+15) F':(1.180083758e+15,-2.547127509e+15) \\
 G:(2.64073598e+15,-9.584372325e+14) G':(2.547127509e+15,1.180083758e+15) \\
 H+:(5.28147196e+15,-1.916874465e+15) H+':(5.094255019e+15,2.360167515e+15) \\
 H-:(7.544527529e-17,1.613446827e-16) H-':(6.131348038e-17,-1.670888938e-16)\\ \\
z:(81.02790609,58.87021973) \\
 l[rec]:(1,0.1) \\
 F:(0.01090170509,0.002924757522) F':(0.00666531837,0.00369511457) \\
 G:(57.24722493,-32.54791916) G':(-42.75162529,10.97359982) \\
 H+:(57.24430017,-32.53701746) H+':(-42.7553204,10.98026514) \\
 H-:(57.25014969,-32.55882087) H-':(-42.74793018,10.9669345)\\ \\l[rec]:(2,0.1) \\
 F:(0.01069308234,0.002972399334) F':(0.006522115638,0.003689236481) \\
 G:(57.93801038,-33.60221658) G':(-43.37569649,11.55133447) \\
 H+:(57.93503798,-33.5915235) H+':(-43.37938573,11.55785658) \\
 H-:(57.94098278,-33.61290967) H-':(-43.37200725,11.54481235)\\ \\l[rec]:(3,0.1) \\
 F:(0.01038725067,0.003043902107) F':(0.006311911265,0.003681809291) \\
 G:(58.95045958,-35.24443848) G':(-44.3055982,12.46205462) \\
 H+:(58.94741568,-35.23405123) H+':(-44.30928001,12.46836653) \\
 H-:(58.95350348,-35.25482573) H-':(-44.30191639,12.45574271)\\ \\
z:(30.94990609,95.25401645) \\
 l[rec]:(1,0.1) \\
 F:(-2.246133078e-15,2.098754042e-15) F':(-5.747597651e-16,2.506104288e-15) \\
 G:(-4.367342673e+13,-1.907186698e+14) G':(1.181536987e+14,1.103266317e+14) \\
 H+:(-4.367342673e+13,-1.907186698e+14) H+':(1.181536987e+14,1.103266317e+14) \\
 H-:(-4.367342673e+13,-1.907186698e+14) H-':(1.181536987e+14,1.103266317e+14)\\ \\
l[rec]:(2,0.1) \\
 F:(-2.280830781e-15,2.034967775e-15) F':(-6.280365023e-16,2.478500727e-15) \\
 G:(-4.826447957e+13,-1.907399944e+14) G':(1.213466147e+14,1.081922962e+14) \\
 H+:(-4.826447957e+13,-1.907399944e+14) H+':(1.213466147e+14,1.081922962e+14) \\
 H-:(-4.826447957e+13,-1.907399944e+14) H-':(1.213466147e+14,1.081922962e+14)\\ \\
l[rec]:(3,0.1) \\
 F:(-2.331318569e-15,1.939356637e-15) F':(-7.06875848e-16,2.436443775e-15) \\
 G:(-5.518815356e+13,-1.904610411e+14) G':(1.260154402e+14,1.047559123e+14) \\
 H+:(-5.518815356e+13,-1.904610411e+14) H+':(1.260154402e+14,1.047559123e+14) \\
 H-:(-5.518815356e+13,-1.904610411e+14) H-':(1.260154402e+14,1.047559123e+14)\\ \\
z:(-30.94990609,95.25401645) \\
 l[rec]:(1,0.1) \\
 F:(-3.696304706e-35,8.374503306e-35) F':(5.116568262e-35,9.162544125e-35) \\
 G:(2.32622983e+33,-4.170545023e+33) G':(2.19801873e+33,4.986604576e+33) \\
 H+:(2.32622983e+33,-4.170545023e+33) H+':(2.19801873e+33,4.986604576e+33) \\
 H-:(2.32622983e+33,-4.170545023e+33) H-':(2.19801873e+33,4.986604576e+33)\\ \\
l[rec]:(2,0.1) \\
 F:(-3.976658859e-35,8.274528706e-35) F':(4.8327284e-35,9.349101023e-35) \\
 G:(2.184541884e+33,-4.231185626e+33) G':(2.351297285e+33,4.898721703e+33) \\
 H+:(2.184541884e+33,-4.231185626e+33) H+':(2.351297285e+33,4.898721703e+33) \\
 H-:(2.184541884e+33,-4.231185626e+33) H-':(2.351297285e+33,4.898721703e+33)\\ \\
l[rec]:(3,0.1) \\
 F:(-4.394887926e-35,8.113966164e-35) F':(4.398972708e-35,9.619612834e-35) \\
 G:(1.968199386e+33,-4.309629699e+33) G':(2.572345582e+33,4.754740671e+33) \\
 H+:(1.968199386e+33,-4.309629699e+33) H+':(2.572345582e+33,4.754740671e+33) \\
 H-:(1.968199386e+33,-4.309629699e+33) H-':(2.572345582e+33,4.754740671e+33)\\ \\
z:(-81.02790609,58.87021973) \\
 l[rec]:(1,0.1) \\
 F:(-2.432130956e-67,3.004725207e-66) F':(3.593950134e-66,1.98955822e-66) \\
 G:(1.065320986e+65,-5.911485322e+64) G':(1.320210869e+64,1.651750319e+65) \\
 H+:(1.065320986e+65,-5.911485322e+64) H+':(1.320210869e+64,1.651750319e+65) \\
 H-:(1.065320986e+65,-5.911485322e+64) H-':(1.320210869e+64,1.651750319e+65)\\ \\
l[rec]:(2,0.1) \\
 F:(-3.576416599e-67,3.029259022e-66) F':(3.559897692e-66,2.145032835e-66) \\
 G:(1.030789257e+65,-6.225220366e+64) G':(1.903651657e+64,1.626497749e+65) \\
 H+:(1.030789257e+65,-6.225220366e+64) H+':(1.903651657e+64,1.626497749e+65) \\
 H-:(1.030789257e+65,-6.225220366e+64) H-':(1.903651657e+64,1.626497749e+65)\\ \\
l[rec]:(3,0.1) \\
 F:(-5.328192626e-67,3.06145691e-66) F':(3.501151213e-66,2.380004106e-66) \\
 G:(9.770243181e+64,-6.655971946e+64) G':(2.74025893e+64,1.583938607e+65) \\
 H+:(9.770243181e+64,-6.655971946e+64) H+':(2.74025893e+64,1.583938607e+65) \\
 H-:(9.770243181e+64,-6.655971946e+64) H-':(2.74025893e+64,1.583938607e+65)\\ \\
z:(-100.156,1.22651674e-14) \\
 l[rec]:(1,0.1) \\
 F:(7.915510206e-34,-4.070932761e-34) F':(3.191449041e-34,1.291346722e-33) \\
 G:(4.180561145e+103,8.128671329e+103) G':(-1.326122108e+104,3.277393326e+103) \\
 H+:(4.180561145e+103,8.128671329e+103) H+':(-1.326122108e+104,3.277393326e+103) \\
 H-:(4.180561145e+103,8.128671329e+103) H-':(-1.326122108e+104,3.277393326e+103)\\ \\
l[rec]:(2,0.1) \\
 F:(7.60252401e-34,-4.273651875e-34) F':(3.593368917e-34,1.252731129e-33) \\
 G:(4.388739394e+103,7.807256555e+103) G':(-1.286466613e+104,3.690136722e+103) \\
 H+:(4.388739394e+103,7.807256555e+103) H+':(-1.286466613e+104,3.690136722e+103) \\
 H-:(4.388739394e+103,7.807256555e+103) H-':(-1.286466613e+104,3.690136722e+103)\\ \\
l[rec]:(3,0.1) \\
 F:(7.127789352e-34,-4.540648434e-34) F':(4.144057432e-34,1.192784614e-33) \\
 G:(4.66292605e+103,7.319737507e+103) G':(-1.224905766e+104,4.255655031e+103) \\
 H+:(4.66292605e+103,7.319737507e+103) H+':(-1.224905766e+104,4.255655031e+103) \\
 H-:(4.66292605e+103,7.319737507e+103) H-':(-1.224905766e+104,4.255655031e+103)\\ \\
z:(-81.02790609,-58.87021973) \\
 l[rec]:(1,0.1) \\
 F:(-23318.74764,-17080.04412) F':(28164.502,-34900.64756) \\
 G:(17080.04412,-23318.74763) G':(34900.64757,28164.50199) \\
 H+:(34160.08824,-46637.49528) H+':(69801.29514,56329.004) \\
 H-:(6.998646405e-06,8.686284079e-06) H-':(1.393963282e-05,-1.022684069e-05)\\ \\
l[rec]:(2,0.1) \\
 F:(-23227.13327,-15830.94943) F':(26224.1359,-34846.45541) \\
 G:(15830.94944,-23227.13327) G':(34846.45543,26224.13589) \\
 H+:(31661.89887,-46454.26654) H+':(69692.91084,52448.27179) \\
 H-:(6.890524871e-06,9.17086489e-06) H-':(1.468279939e-05,-1.002406502e-05)\\ \\
l[rec]:(3,0.1) \\
 F:(-22989.33501,-14034.3731) F':(23425.9144,-34604.63695) \\
 G:(14034.3731,-22989.335) G':(34604.63697,23425.91439) \\
 H+:(28068.7462,-45978.67001) H+':(69209.27392,46851.8288) \\
 H-:(6.703413121e-06,9.918798258e-06) H-':(1.582831668e-05,-9.679629412e-06)\\ \\
z:(-30.94990609,-95.25401645) \\
 l[rec]:(1,0.1) \\
 F:(-3.419604636e+30,-3.206140946e+30) F':(4.148673182e+30,-5.870853625e+30) \\
 G:(3.206140946e+30,-3.419604636e+30) G':(5.870853625e+30,4.148673182e+30) \\
 H+:(6.412281891e+30,-6.839209271e+30) H+':(1.174170725e+31,8.297346364e+30) \\
 H-:(4.01631585e-32,5.68656642e-32) H-':(7.772167843e-32,-7.290658234e-32)\\ \\
l[rec]:(2,0.1) \\
 F:(-3.392855448e+30,-3.007724359e+30) F':(3.853513229e+30,-5.788374417e+30) \\
 G:(3.007724359e+30,-3.392855448e+30) G':(5.788374417e+30,3.853513229e+30) \\
 H+:(6.015448719e+30,-6.785710896e+30) H+':(1.157674883e+31,7.707026457e+30) \\
 H-:(3.986897101e-32,5.991979415e-32) H-':(8.242309638e-32,-7.310395217e-32)\\ \\
l[rec]:(3,0.1) \\
 F:(-3.339981893e+30,-2.724844995e+30) F':(3.435784336e+30,-5.648469464e+30) \\
 G:(2.724844995e+30,-3.339981893e+30) G':(5.648469464e+30,3.435784336e+30) \\
 H+:(5.449689989e+30,-6.679963785e+30) H+':(1.129693893e+31,6.871568672e+30) \\
 H-:(3.932323694e-32,6.468436407e-32) H-':(8.977628494e-32,-7.327940398e-32)\\ \\
z:(30.94990609,-95.25401645) \\
 l[rec]:(1,0.1) \\
 F:(1.125583254e+40,3.548477279e+39) F':(3.759922313e+38,1.698605313e+40) \\
 G:(-3.548477279e+39,1.125583254e+40) G':(-1.698605313e+40,3.759922313e+38) \\
 H+:(-7.096954559e+39,2.251166509e+40) H+':(-3.397210626e+40,7.519844627e+38) \\
 H-:(6.373392558e-43,-2.945348841e-41) H-':(-4.036846684e-41,1.270436304e-41)\\ \\
l[rec]:(2,0.1) \\
 F:(1.092775914e+40,3.208281133e+39) F':(6.877131706e+38,1.638289489e+40) \\
 G:(-3.208281133e+39,1.092775914e+40) G':(-1.638289489e+40,6.877131706e+38) \\
 H+:(-6.416562265e+39,2.185551829e+40) H+':(-3.276578978e+40,1.375426341e+39) \\
 H-:(1.265131507e-42,-3.049938868e-41) H-':(-4.208492895e-41,1.233299224e-41)\\ \\
l[rec]:(3,0.1) \\
 F:(1.043809065e+40,2.740200645e+39) F':(1.099037848e+39,1.550082818e+40) \\
 G:(-2.740200645e+39,1.043809065e+40) G':(-1.550082818e+40,1.099037848e+39) \\
 H+:(-5.480401289e+39,2.087618131e+40) H+':(-3.100165636e+40,2.198075696e+39) \\
 H-:(2.26211583e-42,-3.213069851e-41) H-':(-4.477068961e-41,1.172939493e-41)\\ \\
z:(81.02790609,-58.87021973) \\
 l[rec]:(1,0.1) \\
 F:(-2.579395538e+32,7.380968215e+32) F':(-9.701448491e+32,1.926734514e+32) \\
 G:(-7.380968215e+32,-2.579395538e+32) G':(-1.926734514e+32,-9.701448491e+32) \\
 H+:(-1.476193643e+33,-5.158791075e+32) H+':(-3.853469027e+32,-1.940289698e+33) \\
 H-:(-4.963907579e-34,-9.779175603e-35) H-':(2.098665903e-34,6.035174253e-34)\\ \\
l[rec]:(2,0.1) \\
 F:(-2.416251979e+32,7.159184131e+32) F':(-9.355630991e+32,1.963184002e+32) \\
 G:(-7.159184131e+32,-2.416251979e+32) G':(-1.963184002e+32,-9.355630991e+32) \\
 H+:(-1.431836826e+33,-4.832503957e+32) H+':(-3.926368004e+32,-1.871126198e+33) \\
 H-:(-5.124854379e-34,-1.067175763e-34) H-':(2.104701743e-34,6.26792112e-34)\\ \\
l[rec]:(3,0.1) \\
 F:(-2.191409619e+32,6.833977857e+32) F':(-8.857660418e+32,2.000904331e+32) \\
 G:(-6.833977857e+32,-2.191409619e+32) G':(-2.000904331e+32,-8.857660418e+32) \\
 H+:(-1.366795571e+33,-4.382819238e+32) H+':(-4.001808662e+32,-1.771532084e+33) \\
 H-:(-5.377076882e-34,-1.205969114e-34) H-':(2.115384163e-34,6.632005233e-34)\\ \\ \\
Recurrence relations results for a single z. \\
z:(100.156,0) \\
 l[rec]:(1,0.1) \\
 F:(-1.021072923e+15,-2.836755456e+15) F':(1.275057299e+15,-2.729507771e+15) \\
 G:(2.836755456e+15,-1.021072923e+15) G':(2.729507771e+15,1.275057299e+15) \\
 H+:(5.673510913e+15,-2.042145845e+15) H+':(5.459015542e+15,2.550114597e+15) \\
 H-:(7.0774288e-17,1.501204734e-16) H-':(5.67178338e-17,-1.558437769e-16)\\ \\
l[rec]:(2,0.1) \\
 F:(-9.963131598e+14,-2.756374147e+15) F':(1.235449857e+15,-2.655381893e+15) \\
 G:(2.756374147e+15,-9.963131598e+14) G':(2.655381893e+15,1.235449857e+15) \\
 H+:(5.512748294e+15,-1.99262632e+15) H+':(5.310763785e+15,2.470899715e+15) \\
 H-:(7.256451117e-17,1.54534022e-16) H-':(5.855852326e-17,-1.602324976e-16)\\ \\
l[rec]:(3,0.1) \\
 F:(-9.584372325e+14,-2.64073598e+15) F':(1.180083758e+15,-2.547127509e+15) \\
 G:(2.64073598e+15,-9.584372325e+14) G':(2.547127509e+15,1.180083758e+15) \\
 H+:(5.28147196e+15,-1.916874465e+15) H+':(5.094255019e+15,2.360167515e+15) \\
 H-:(7.544527529e-17,1.613446827e-16) H-':(6.131348038e-17,-1.670888938e-16)
}

\newpage

\begin{table}[htbp]
\begin{center}
\caption{Number of iterations $n$ needed for the continued fraction $h^{+}(z)$ to converge up to $\epsilon = 10^{-10}$ with Lentz method.
The set of parameters is chosen here as $\ell = 0$, $\eta = 10$, and $z = x - 2i$, with $x$ varying between 0 and 1.}
\vspace*{1cm}
\begin{tabular}{|c|c|} \hline
$x$ & $n$ \\ \hline
1 & 485 \\
0.5 & 1,679 \\
0.1 & 35,984 \\
0.05 & 137,922 \\
0.01 & 3,146,899 \\
0.005 & 12,107,924 \\ \hline
\end{tabular}
\label{table_slow_cv}
\end{center}
\end{table}

\vspace*{1cm}

\begin{table}[htbp]
\begin{center}
\caption{Comparison of the Coulomb wave functions $F_{\ell,\eta}(z)$ and $G_{\ell,\eta}(z)$ and derivatives
calculated with direct and first-order expansion methods. The second and third columns show respectively the real and imaginary parts
of the Coulomb wave functions obtained with the direct method.
The fourth and fifth column provides the relative difference of the same real and imaginary parts 
with those calculated with the first-order expansion respectively.
The used set of parameters is $\ell = i10^{-5}$, $\eta = 10 + i10^{-5}$, and $z = 0.1 + i10^{-6}$.}
\vspace*{1cm}
\begin{tabular}{|c|c|c|c|c|} \hline
      & Direct (real part)     & Direct (im.~part)       & Rel.~diff.~(real part) & Rel.~diff.~(im.~part) \\ \hline
$F$   & 4.306$\cdot 10^{-14}$ & --9.033$\cdot 10^{-19}$ & 1.481$\cdot 10^{-10}$ & 6.819$\cdot 10^{-10}$ \\
$F'$  & 7.635$\cdot 10^{-13}$ & --1.861$\cdot 10^{-17}$ & 2.507$\cdot 10^{-10}$ & 5.467$\cdot 10^{-10}$ \\
$G$   & $\!\!\!\!$7.787$\cdot 10^{11}$  &  $\!\!\!$1.842$\cdot 10^{7}$  & 3.299$\cdot 10^{-10}$ & 1.053$\cdot 10^{-9}$ \\ 
$G'$  & $\!\!\!\!\!\!\!$--9.416$\cdot 10^{12}$ & $\!\!\!\!\!\!\!$--2.076$\cdot 10^{8}$   & 3.086$\cdot 10^{-10}$ & 2.849$\cdot 10^{-10}$\\ \hline
\end{tabular}
\label{tab_first_order_exp}
\end{center}
\end{table}
 
\begin{table}[htbp]
\begin{center}
\caption{Energies and widths of the $2s_{1/2}$ proton state of the Hamiltonian $h$ of Eq.~(\ref{h_proton_emitter})
as a function of the depth of the Woods-Saxon potential $V_0$ given in MeV. $\gamma$ denotes the width obtained by direct integration of $h$,
and $\gamma_{c}$ is the width obtained by the approximate current formula of Eq.~(\ref{gamma_app}).
Energies are given in MeV and widths in keV.}
\vspace*{1cm}
\begin{tabular}{|c|c|c|c|} \hline
$V_0$ (MeV) & $E$ (MeV) & $\gamma$ (keV) & $\gamma_{c}$ (keV) \\ \hline 
50 & 4.510 & 6.188$\cdot 10^{-1}$ & 6.188$\cdot 10^{-1}$ \\
51 & 3.847 & 6.134$\cdot 10^{-2}$ & 6.134$\cdot 10^{-2}$ \\
52 & 3.168 & 2.597$\cdot 10^{-3}$ & 2.597$\cdot 10^{-3}$ \\
53 & 2.477 & 2.664$\cdot 10^{-5}$ & 2.664$\cdot 10^{-5}$ \\
54 & 1.773 & 1.777$\cdot 10^{-8}$ & 1.777$\cdot 10^{-8}$ \\
55 & 1.060 & 1.260$\cdot 10^{-14}$ & 1.260$\cdot 10^{-14}$ \\
56 & 0.336 & 7.738$\cdot 10^{-36}$ & 7.738$\cdot 10^{-36}$ \\
56.4 & 4.458$\cdot 10^{-2}$ & 4.956$\cdot 10^{-121}$ & 4.956$\cdot 10^{-121}$ \\ 
56.46 & 6.949$\cdot 10^{-4}$ & 0 & 0 \\ \hline
\end{tabular}
\label{table_proton_emitter}
\end{center}
\end{table}

\newpage

\begin{thebibliography}{10}
 
\bibitem{Abramowitz_Coulomb} M.~Abramowitz, Chap.~14 ``Coulomb Wave Functions'', Handbook of Mathematical Functions, 
edited by M.~Abramowitz and I.A.~Stegun, National Bureau of Standards, Applied Mathematics Series - 55 (1972).
\bibitem{Humblet} J.~Humblet, Nucl. Phys. 50 (1964) 1; Ann. Phys. 155 (1984) 461.
\bibitem{Kolbig} K.S.~K{\"o}lbig, Comp. Phys. Comm. 4 (1972) 221.
\bibitem{Gamow} G.A.~Gamow, Zs. f. Phys. 51 (1928) 204; 52 (1928) 510.
\bibitem{Regge_pole_ex} A.~Amaha et al., Phys. Rev. A 45  (1992) 1596.
\bibitem{Seaton} M.J.~Seaton, Comp. Phys. Comm. 146 (2002) 225.
\bibitem{Barnett_real} A.R.~Barnett, J. Comput. Phys. 46 (1982) 171.
\bibitem{Tamura} T.~Tamura and F.~Rybicki, Comp. Phys. Comm. 1 (1969) 25.
\bibitem{Takemasa} T.~Takemasa, T.~Tamura and H.H.~Wolter, Comp. Phys. Comm. 17 (1979) 351.
\bibitem{Noble} C.J.~Noble, Comp. Phys. Comm. 159 (2004) 55.
\bibitem{Thompson} I.J.~Thompson and A.R.~Barnett, Comp. Phys. Comm. 36 (1985) 363; J. Comput. Phys. 64 (1986) 490.
\bibitem{Cernlib} CERNLIB library,  C309: Coulomb Wave, Bessel, and Spherical Bessel Functions for Complex Argument(s) and Order \\
                  http://wwwasdoc.web.cern.ch/wwwasdoc/shortwrupsdir/c309/top.html
\bibitem{Kruppa_comm} A.T.~Kruppa, private communication.
\bibitem{Numerical_Recipes} W.H.~Press, S.A.~Teukolsky, W.T.~Vetterling and B.P.~Flannery, Numerical Recipes in C, Cambridge University Press 1988-1992.
\bibitem{Todd} J.~Todd, Survey of Numerical Analysis, McGraw-Hill Book Company, Inc., New York 1962.
\bibitem{Gautschi} W.~Gautschi, Math. Comp. 31 (1977) 994.
\bibitem{Dzieciol} A.~Dzieciol, S.~Yngve and P.O.~Fr{\"o}man, J. Math. Phys. 40 (1999) 6145.
\bibitem{Kruppa_PRL} A.T.~Kruppa, B.~Barmore, W.~Nazarewicz and T.~Vertse, Phys. Rev. Lett. 19 (2000) 4549.
\bibitem{Kruppa_proc} A.T.~Kruppa, N.~Michel and W.~Nazarewicz,
in Proceedings of the International Conference on Nuclear Physics,
Large and Small: Microscopic Studies of Collective Phenomena, \\ 
presented by W.~Nazarewicz at Cocoyoc, Morelos, Mexico, April 19-22, 2004 \\ 
Eds: Bijker, R. et al. New York, AIP (AIP Conference Proceedings 726) (2004) 7.
\end{thebibliography}
\end{document}